# Radiation damage free ghost diffraction with atomic resolution

Zheng Li[1,2,3,*], Nikita Medvedev[1,4], Henry N. Chapman[1,2,5], and Yanhua Shih[6,*]

[*]zheng.li@desy.de
[1]Center for Free-Electron Laser Science, DESY, Notkestraße 85, D-22607 Hamburg, Germany
[2]Hamburg Centre for Ultrafast Imaging, Luruper Chaussee 149, D-22761 Hamburg, Germany
[3]SLAC National Accelerator Laboratory, Menlo Park, California 94025, USA
[4]Institute of Physics and Institute of Plasma Physics, Academy of Science of Czech Republic, Na Slovance 1999/2, 18221 Prague 8, Czech Republic
[5]Department of Physics, University of Hamburg, Jungiusstraße 9, D-20355 Hamburg, Germany
[*]shih@umbc.edu
[6]Department of Physics, University of Maryland, Baltimore County, Baltimore, Maryland 21250, USA

## ABSTRACT

The X-ray free electron lasers (XFEL) can enable diffractive structural determination of protein nanocrystals or single molecules that are too small and radiation-sensitive for conventional X-ray analysis, because ultrashort X-ray pulses from XFEL can reduce the radiation damage. However the electronic form factor may be modified during the ultrashort X-ray pulse due to photoionization and electron cascade caused by the intense X-ray pulse. For general X-ray imaging techniques, the minimization of the effects of radiation damage is of major concern to ensure reliable reconstruction of molecular structure. Here we show that radiation damage free diffraction can be achieved with atomic spatial resolution by using X-ray parametric down-conversion (XPDC) and ghost diffraction with entangled photons of X-ray and optical frequencies. We show that the formation of the diffraction patterns satisfies a condition analogous to the Bragg equation, with a resolution that can be as fine as the crystal lattice length scale of several Ångstrom. Since the samples are illuminated by optical photons of low energy, they can be free of radiation damage.

## Introduction

The advent of femtosecond X-ray free electron lasers (XFEL) has enabled diffractive structural determination of protein crystals or single molecules that are too small and radiation-sensitive for conventional X-ray analysis[1–3]. Using X-ray pulses of $\sim$ 10fs, sufficient diffraction signals could be collected before significant changes occur in the sample[1]. Nevertheless, the electronic form factor could be modified due to photoionization and electron cascades caused by the intense X-ray pulse[4,5]. For general X-ray imaging techniques, minimizing the effects of radiation damage is of major concern to guarantee a reliable reconstruction of the structure. Here we show that a radiation damage free diffraction is achievable with an atomic spatial resolution by using X-ray parametric down-conversion (XPDC)[6–8] and ghost diffraction with entangled photons of X-ray and optical frequencies. An intense hard X-ray pulse generated by XFEL is used to pump a nonlinear medium, and is down-converted to two-photon pairs that consist of an X-ray and an optical photon with wavelengths $\lambda_X$ and $\lambda_o$, respectively. The optical photons are sent to illuminate the sample crystals or molecules, and the reflected photons are collected with a bucket detector $D_A$. The X-ray photons entangled with the optical photons travel a certain distance, and are captured by a pixel photon counting detector $D_B$. Based on the basic principle of ghost imaging[9–14], the output pulses of the bucket detector $D_A$ and the pixel detector $D_B$ are sent to a coinicidence circuit with certain time gate for counting the joint-detection of the two-photon pairs.

We show here that the X-ray photons can form diffraction patterns on pixel detectors. We illustrate that the formation of the diffraction patterns satisfies a condition analogous to the Bragg equation with a resolution that could be as fine as the lattice length scale of several ångstrom. Because the samples are illuminated with the optical photons of low energy, they can be free of radiation damage. The ultrabright intensity of XFEL could be crucial for realization of the proposed scheme to ensure sufficient two-photon flux and signal strength. Since the diffraction pattern formation is based on photon counting, the requirement for signal intensity is benign, we also show that the analogous Bragg condition can be satisfied with feasible experimental parameters.

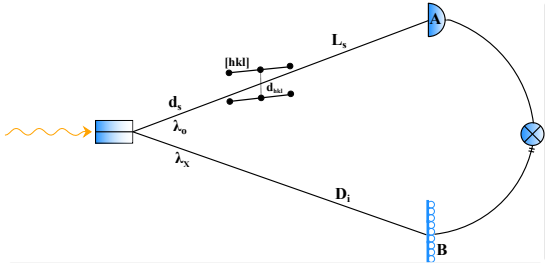

**Figure 1.** Proposed layout for two-color two-photon ghost diffraction using entangled X-ray and optical photon pairs from XPDC. The optical photons with $\lambda_o$ scatter off lattice planes with Miller index $[hkl]$ and inter-plane distance $d_{hkl}$, and are detected by the bucket detector $D_A$. The distances from the crystal plane to that XPDC output plane and the plane of bucket detector is $d_s$ and $L_s$ respectively. The entangled X-ray photons with $\lambda_X$ travel a distance of $D_i$ to the pixelated detector $D_B$. Correlation measurement is carried out for the detectors $D_A$ and $D_B$.

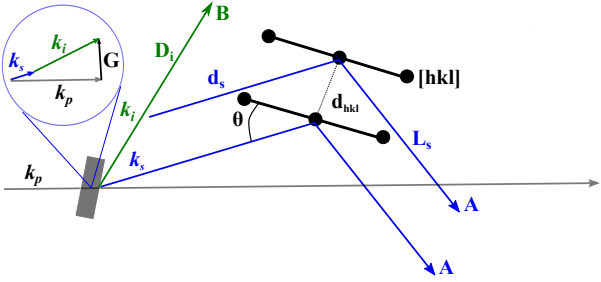

**Figure 2.** Sketch of the proposed experiment. The lattice plane of the sample forms an angle of $\theta$ with the incident optical photon in the arm A. The relative angle of the X-ray photon (arm B) with the pump photon is determined from the phase matching diagram, $\vec{k}_s + \vec{k}_i = \vec{k}_p + \vec{G}$, where $\vec{G}$ is a reciprocal lattice vector orthogonal to certain atomic planes of a nonlinear crystal; $\vec{k}_s$, $\vec{k}_i$ and $\vec{k}_p$ are the wave vectors of the signal, idler and pump fields. And $z_s = d_s + L_s$, $z_i = D_i$ are the optical path lengths of the signal and idler photons.

## Results

An optical beam and an X-ray beam from XPDC are directed to illuminate a crystal in the optical arm and form diffraction patterns in the X-ray arm (Fig. 1). Without loss of generality, we assume that the signal photon has an optical wavelength of $\lambda_s = \lambda_o$, and the idler photon has an X-ray wavelength of $\lambda_i = \lambda_X$. From the detailed derivation presented in Section S1 of Supplementary Information (Eq. S1–S24)[15], as the optical photons scatter off lattice planes of inter-plane distance $d_{hkl} = d$, the condition for the X-ray photons to form peaks in the diffraction pattern is

$$2\tilde{m}d\sin\theta = n\lambda_s, \qquad (1)$$

for an integer $n$. $\theta$ is the reflection angle of the optical photon from the lattice planes of a Miller index $[hkl]$ (Fig. 2), and $\vec{\rho}_B$ is a vector on the plane of the pixel detector. The magnification factor $\tilde{m}$ is found to be (Eq. S23)[15]

$$\tilde{m} = 1 - \frac{\left|\vec{\rho}_B - \vec{d}\right|^2}{(\alpha+1)\left(\frac{\lambda_i}{\lambda_s}\right)^2 D_i^2}, \qquad (2)$$

where $\alpha$ is defined as $\frac{d_s}{D_i} = \alpha \frac{\lambda_i}{\lambda_s}$. $\tilde{m}$ guarantees that the Bragg condition (Eq. 1) can be satisfied in the case $d \ll \lambda_s$.

The two-photon coincidence counting rate of the bucket detector A and the pixel detector B can be written as[10,12],

$$R_c(\vec{\rho}_B) = \frac{1}{T}\int dt_A dt_B S(t_B, t_A)\sigma_B \text{tr}_{\vec{\rho}_A}\left[\hat{E}_A^{(-)}\hat{E}_B^{(-)}\hat{E}_B^{(+)}\hat{E}_A^{(+)}\hat{\rho}\right], \qquad (3)$$

where $S(t_B, t_A)$ is the coincidence time function that vanishes unless $0 \leq t_B - t_A \leq T$, and describes the finite time gate of a joint-detection of two-photon pair[10]. $\vec{\rho}_A$ is on the plane of the bucket detector A with an area $\sigma_A$, and $\sigma_B$ is the area of the pixel detector B. $\rho$ is the density matrix of the two-photon state on the output plane of the nonlinear crystal. $E_j^{(+)}$, $j = A, B$ is the positive frequency part of the photon field, and $E_j^{(-)} = (E_j^{(+)})^\dagger$. $\text{tr}_{\vec{\rho}_A}[\cdots]$ denotes the trace, i.e. coherent summation over $\vec{\rho}_A$. $R_c T$ gives the pixelwise number of counted photons in one joint measurement. The photon fields at the plane of the bucket detector A and the pixel detector B, $\hat{E}_A^{(+)}$ and $\hat{E}_B^{(+)}$ can be expressed as

$$\begin{aligned}\hat{E}_A^{(+)} &= \sum_{\vec{k}_s} g(\vec{\kappa}_s, \omega_s, \vec{\rho}_A, z_s) \hat{a}_s(\vec{k}_s) e^{-i\omega_s t_A} \\ \hat{E}_B^{(+)} &= \sum_{\vec{k}_i} g(\vec{\kappa}_i, \omega_i, \vec{\rho}_B, z_i) \hat{a}_i(\vec{k}_i) e^{-i\omega_i t_B},\end{aligned} \quad (4)$$

where $z_s = d_s + L_s$ and $z_i = D_i$ are the full optical path lengths for the signal and the idler photon, $\vec{\kappa}$ is the transverse momentum of the photon field with $\vec{k} = \sqrt{k^2 - \kappa^2} \hat{e}_z + \vec{\kappa}$. $\hat{a}_p^\dagger(\vec{k})$ and $\hat{a}_p(\vec{k})$ are creation and annihilation operators of the signal and the idler photon field in a specific mode at the output plane of the nonlinear crystal, with the commutator relation, $\left[\hat{a}_p(\vec{k}), \hat{a}_q^\dagger(\vec{k}')\right] = \delta_{p,q} \delta_{\omega,\omega'} \delta_{\vec{\kappa},\vec{\kappa}'}$. $g(\vec{\kappa}, \omega, \vec{\rho}, z)$ is the Green's function for a specific mode of the photon field. Assume two atoms in two lattice planes of Miller index $[hkl]$ with distance $d = d_{hkl}$ are in a plane $a$, $\vec{\rho}_a$ is a vector in this plane, and the photon-atom scattering amplitude is $t(\vec{\rho}_a)$, the photon fields $\hat{E}_A^{(+)}$ and $\hat{E}_B^{(+)}$ at the planes of detectors can be expressed as,

$$\begin{aligned}\hat{E}_A^{(+)} &= \sum_{\vec{k}_s} \int d^2 \vec{\rho}_{s'} \int d^2 \vec{\rho}_a \frac{-i k_s}{2\pi d_s} e^{i k_s d_s} e^{i \vec{\kappa}_s \cdot \vec{\rho}_{s'}} e^{i \frac{k_s}{2 d_s} |\vec{\rho}_a - \vec{\rho}_{s'}|^2} \\ &\quad \times t(\vec{\rho}_a) \frac{-i k_s}{2\pi L_s} e^{i k_s L_s} e^{i \frac{k_s}{2 L_s} |\vec{\rho}_A - \vec{\rho}_a|^2} e^{-i \omega_s t_A} \hat{a}_s(\vec{k}_s) \\ \hat{E}_B^{(+)} &= \sum_{\vec{k}_i} \int d^2 \vec{\rho}_s \frac{-i k_i}{2\pi D_i} e^{i k_i D_i} e^{i \vec{\kappa}_i \cdot \vec{\rho}_s} e^{i \frac{k_i}{2 D_i} |\vec{\rho}_s - \vec{\rho}_B|^2} e^{-i \omega_i t_B} \hat{a}_i(\vec{k}_i).\end{aligned} \quad (5)$$

Using first-order perturbation theory, the two-photon amplitude from an XPDC process is shown to be[15]

$$\langle 0 | \hat{a}_s(\vec{k}_s) \hat{a}_i(\vec{k}_i) | \Psi \rangle = -i(2\pi)^3 \gamma \delta(v_s + v_i) \delta(\vec{\kappa}_s + \vec{\kappa}_i) \text{sinc}(v_s D_{si} \frac{L}{2}), \quad (6)$$

where $|\Psi\rangle$ is the two-photon state vector, $\gamma = \frac{\chi^{(2)} E_p L}{2 U_s U_i} \sqrt{\frac{\Omega_s \Omega_i T_s T_i c^2}{n_s n_i}}$, $\chi^{(2)}$ is the second-order susceptibility of the nonlinear crystal, $U_j$ is the group velocity of the signal and the idler photon inside the nonlinear crystal of length $L$, $T_j$ are their transmission coefficients, and $E_p$ is the strength of the pump field. $\Omega_s$ and $\Omega_i$ are the frequencies of the signal and the idler photon that satisfy the phase matching condition, $\omega_p n_p(\omega_p) - \Omega_s n_s(\Omega_s) - \Omega_i n_i(\Omega_i) = 0$. Provided that the Bragg equation (Eq. 1) and the phase matching condition are satisfied, diffraction patterns can be formed on the X-ray photon arm from the coincidence photon count in a joint measurement of the bucket detector $D_A$ and the pixel detector $D_B$ such hat the reflection angle $\theta$ is measured with the detector $D_A$, and the corresponding intensity is measured with the detector $D_B$ from coincidence counts on pixels of a given radius $|\vec{\rho}_B|$. For the schematic configuration in Fig. 1, the Bragg peaks are manifested as modulation of the coincident counting rate at $\vec{\rho}_B$ in the plane of the pixel detector with a background[15]

$$R_c(\vec{\rho}_B) = \frac{1}{T} \int dt_A dt_B S(t_B, t_A) \sigma_B \left(\frac{\gamma \psi_0}{\pi \lambda_s R_s}\right)^2 \left|\int d\nu_s e^{i\nu_s \tau_{BA}} \text{sinc}(\nu_s D_{si} \frac{L}{2})\right|^2 \cos^2\left[\frac{2\pi}{\lambda_s R_s} \vec{d} \cdot (\vec{\rho}_B - \frac{\vec{d}}{2})\right], \quad (7)$$

where $\gamma$ and $D_{si}$ are parameters of the XPDC process[15], $D_{si} = \frac{1}{U_s} - \frac{1}{U_i}$, and $\tau_{BA} = \tau_B - \tau_A$ with $\tau_i = t_i - \frac{r_i}{c}$, $i = A, B$. $\psi_0$ is the scattering amplitude of an atom in the lattice with the optical photons. And $R_s = \frac{\lambda_i}{\lambda_s} D_i + d_s$ is the effective optical path length from the sample to the pixel detector B. We assume the phase matching condition to be $\Omega_s n_s(\Omega_s) + \Omega_i n_i(\Omega_i) = c k_p$, ignoring $\vec{G}$ for simplicity, which can be restored for a given experimental configuration. The frequency of the signal field is $\omega_s = \Omega_s + \nu_s$, with $\nu_s$ characterizing the deviation from the central frequency $\Omega_s$ of the phase matching condition, and $\nu_s = -\nu_i$.

For realistic experimental consideration, we assume an X-ray beam of 3.1 keV ($\lambda_i = 4\text{Å}$) and an optical beam of 3.1 eV ($\lambda_s = 4000\text{Å}$). Nonlinear X-ray process with energy difference of the two-photon on such scale was experimentally demonstrated[6, 16]. Assume the sample is placed at a distance of 1 cm from the XPDC source, the reflection angle of optical photon is measured on detector $D_A$, and the x-ray photon in the two-photon pair is measured by the pixels with radius of 1 m on the pixel detector $D_B$ placed at a distance of 10 m. In the XFEL nanocrystal diffraction experiment, virtual powder diffraction patterns are formed for various reflection angles. The singles photons from the XPDC process have a certain angular spread[17]. It is practically advantageous to determine the actual reflection angle $\theta$ by a pixel detector $D_A$, although in an idealized setup the reflection angle can be uniquely determined by the geometrical configuration of the photon source and a single crystal, for which a bucket detector $D_A$ is sufficient. The Bragg equation gives the ghost diffraction pattern in Fig. 3. The resolution requirement can be satisfied with the state-of-the-art pixel pnCCD detector that can achieve $75\mu$m pixel pitch[1]. The requirement to the spectral and angular resolution of the diffractometer determined by $\frac{\Delta d}{d} + \cot\theta \Delta\theta = \frac{\Delta\lambda_s}{\lambda_s}$, which is on the same scale of the conventional Laue diffraction. For nanocrystals with the size $L_c$, the analogous Scherrer equation is found to be (see Section S3 of Supplementary Information, Eq. S40–S43)[15]

$$B = \frac{2\lambda_s}{L_c \cos\theta \tilde{m}}, \tag{8}$$

which determines the width $B$ of Bragg peaks. As shown Fig. 3, the width of Bragg peaks is on the same scale of Laue diffraction and is resolvable by the state-of-the-art diffractometer. The optical photon energy of 3.1 eV may fall below band gap of the crystal, thus significant absorption can be avoided. The proposed two-color entangled ghost diffraction approach could eventually achieve atomic resolution without radiation damage.

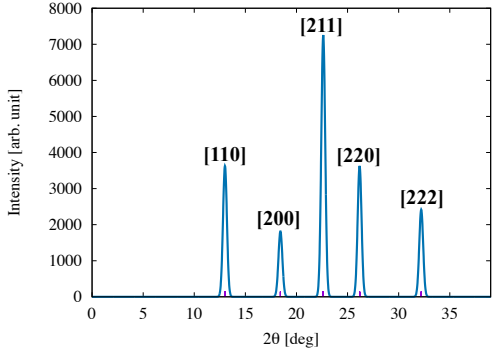

**Figure 3.** Simulated two-color two-photon ghost diffraction from a body centered cubic (bcc) crystal with lattice constant $a = b = c = 4Å$. Optical photons of 3.1 eV are used to illuminate the sample crystal. Miller indices are labelled for the Bragg peaks. The broadening of Bragg peaks is determined by the Scherrer equation (Eq. 8) for a nanocrystal of 100 nm size.

## Discussion

The physical mechanism of the two-color two-photon ghost diffraction can be understood from the particle nature of the optical and X-ray photons and their position-momentum entanglement. The optical photon of frequency $\omega$ scatters with the atoms in a nanocrystal or a molecule and experiences momentum transfer $\vec{Q}$, with frequency independent Thomson scattering cross section $\frac{d\sigma(\vec{Q})}{d\Omega_{\text{th}}} = \frac{r_e^2}{2}(1+\cos^2\theta)\left|f(\vec{Q})\right|^2$, provided the optical photons are unpolarized, where $r_e$ is the classical electron radius $r_e \simeq 2.82$ fm, and $f(\vec{Q})$ is the form factor. The elastic Rayleigh scattering of optical photon with the cross section $\frac{d\sigma(\vec{Q})}{d\Omega} \simeq \frac{\omega^4}{(\omega^2-\omega_0^2)^2}\frac{r_e^2}{2}(1+\cos^2\theta) \simeq r_e^2|f'(\omega)+if''(\omega)|^2$ is equivalent to the dispersive corrections of form factor in the X-ray regime (see explanation in Section S5 of Supplementary Information)[15], which can be used for the phase problem in crystallography[18]. For photons as particles, the probability of reflection from an atom is in fact on the similar orders of magnitude for the optical and X-ray photons. Without considering the molecular form factor of $f(\vec{Q})$, we could model the form factor of a crystal as periodic distributions of point scatterers as $f(\vec{Q}) = -r_e \delta(\vec{Q} - 2\pi\vec{G}_{hkl})$, where $\vec{G}_{hkl}$ is the reciprocal vector with $|\vec{G}_{hkl}| = \frac{1}{d_{hkl}}$. However, for the scattering of an optical photon an atoms, the momentum transfer $\vec{Q} = 2\frac{\omega}{c}\sin\theta(-\cos\phi\cos\theta, -\sin\phi\cos\theta, \sin\theta)$, which is proportional to the incident optical photon energy, is too small to form interference patterns at non-imaginary reflection angle because $\vec{Q} \ll \vec{G}_{hkl}$. Thus, Laue diffraction requires X-ray photons of wavelength $\lambda < 2d$.

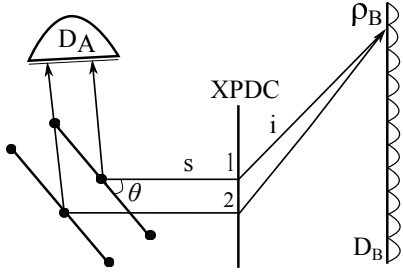

**Figure 4.** Unfolded two-photon diagram of the proposed experimental setup. The optical paths of the x-ray and optical photons are effectively concatenated at position 1 or 2 on the XPDC plane, because $\Delta(\vec{k}_s + \vec{k}_i) = 0$ and $\Delta(\vec{\rho}_s - \vec{\rho}_i) = 0$ are fulfilled at the same time for the two position-momentum entangled particles[19].

The momentum transfer can be effectively magnified in the two-photon ghost diffraction from $\vec{Q}$ to $\tilde{m}\vec{Q}$. The mechanism of momentum transfer magnification can be understood by a simple quantum model of the "unfolded" two-photon ghost diffraction[20] (see Fig. 4). In this simplified model, we do not consider the transverse momenta of the down converted photons as in the exact treatment presented in the previous section. The photon fields on detectors $D_A$ and $D_B$ can be written in this case as

$$\begin{aligned}\hat{E}_A^{(+)} &= \hat{a}_{1s}e^{ik_s r_{A1}} + \hat{a}_{2s}e^{ik_s r_{A2}} \\ \hat{E}_B^{(+)} &= \hat{a}_{1i}e^{ik_i r_{B1}} + \hat{a}_{2i}e^{ik_i r_{B2}},\end{aligned} \qquad (9)$$

where $\hat{a}_{lp}$, $l = 1, 2$, $p = s, i$ are the annihilation operators of signal and idler photons at position 1, 2 in Fig. 4. The two-photon state $|\Psi\rangle$ can be expressed as

$$|\Psi\rangle = |0\rangle + \varepsilon[\hat{a}_{1s}^\dagger \hat{a}_{1i}^\dagger e^{i\phi_1} + \hat{a}_{2s}^\dagger \hat{a}_{2i}^\dagger e^{i\phi_2}]|0\rangle, \qquad (10)$$

where $\varepsilon$ is the two-photon amplitude, and we can assume $\phi_1 = \phi_2$ since the pump beam from an XFEL or a synchrotron must be transversely coherent at position 1 and 2. The two-photon interference is formed due to the uncertainty in the birth place of the two-photon pair (either at position 1 or 2 in Fig. 4), and the property of the two-photon product state that allows us to have the uncertainty relation $\Delta(\vec{k}_s + \vec{k}_i) = 0$ and $\Delta(\vec{\rho}_s - \vec{\rho}_i) = 0$ at the same time[19], which guarantees the operation to concatenate the optical paths of the X-ray and optical photons exactly at the point of their birth position. It is straightforward to find the second-order coherence function $G_{AB} \propto \varepsilon^2 \cos^2\left[\frac{\pi}{\lambda_s}(r_{A1} - r_{A2}) + \frac{\pi}{\lambda_i}(r_{B1} - r_{B2})\right]$, and Bragg peak is formed under condition

$$2d\sin\theta + \frac{\lambda_s}{\lambda_i}(r_{B1} - r_{B2}) = n\lambda_s, \qquad (11)$$

In Eq. 11, the Bragg condition can be satisfied despite $d \ll \lambda_s$, because the optical path length difference can be compensated by that of the arm of idler photons magnified by a factor $\frac{\lambda_s}{\lambda_i}$ of $\sim 10^3$. As a result, we have a qualitative explanation to the origin of the magnification factor $\tilde{m}$ that appears in Eq. 1, and equivalently to the effective form factor $f_{\text{eff}}(\vec{Q}) = -r_e \delta(\tilde{m}\vec{Q} - 2\pi\vec{G}_{hkl})$, which is able to produce diffraction patterns for reciprocal vectors on the 1/Å scale. It is important to notice that the photon behaves as point particle in the elastic scattering, and its wavelength is irrelevant to the effective size of the photon, thus the optical photon does not smear out the crystal structure with period that is much shorter than its wavelength. In fact the wavelength and the size of a quantum object reflect its wave-like and particle-like natures respectively, and should be distinguished from each other. It is known that the intensity of Bragg peaks should be reduced due to the dynamic structural factor $S(\vec{Q}, \omega_s)$[21] by a factor of $\exp\left\{-\frac{1}{2}\left\langle\left[\vec{Q} \cdot (\vec{u}_{\vec{R}}(t) - \vec{u}_{\vec{0}}(0))\right]^2\right\rangle_T\right\}$, where $\vec{u}_{\vec{R}}(t)$ and $\vec{u}_{\vec{0}}(0)$ are the displacement vectors around $\vec{0}$ and $\vec{R}$, and $\vec{R}$ is the lattice vector, $\langle\cdots\rangle_T$ denotes the thermal expectation value. Due to the small momentum transfer $\vec{Q}$, the reduction of Bragg peak intensity is substantially weaker than that of Laue diffraction.

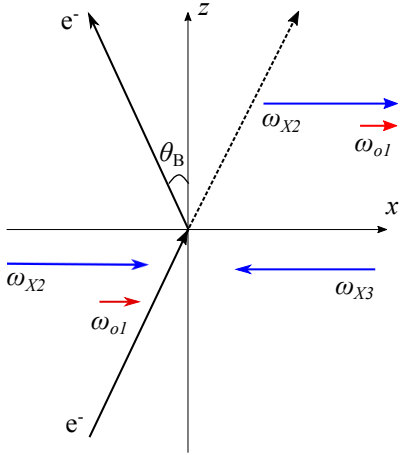

**Figure 5.** Sketch of the seeded Kapitza-Dirac-like XPDC process using single electron as the nonlinear medium. An X-ray photon of $\omega_{X3}$ is down converted to two photons of frequencies $\omega_{o1}$ and $\omega_{X2}$, as the electron is deflected with Bragg angle $\theta_B$ in coincidence. For the use of ghost diffraction, the down converted photon pair can be then spatially separated by a metal foil that reflects the optical photon. The collinear geometry of $k_{X3}$ and $(k_{o1}, k_{X2})$ can be loosened for more convenient experimental setup, provided the phase matching condition is satisfied.

Because the frequency of the optical photon is much lower than the X-ray photon, $\omega_o \ll \omega_X$, the experiment would suffer from strong absorption from photon-bound electron scattering. Thus the optical photons in the vacuum ultraviolet (VUV) regime is not favorable for the proposed ghost diffraction. The entanlged ghost diffraction scheme using optical photons to interact with the crystal could also potentially suffer from skin effect when the samples are good conductors like copper, which has ∼3 nm skin depth for optical photons in the visible and ultraviolet regime, and optical photons can only penetrate several uppermost lattice planes. In this sense, the proposed ghost diffraction scheme is suitable to diffractive structural determination of insulators, some semiconductors and proteins, which should have poor conductivity. For good conductors, our method is limited to the cases of thin and homogeneous samples, or the studies of surfaces.

In order to achieve sufficient photon counting rate, it is crucially important to have a two-photon source with high flux using X-ray quantum optical techniques[22]. Because the XPDC and the sum-frequency generation (SFG) processes[6] are subject to the same nonlinear susceptibility, we can expect the feasibility of two-photon production with similar $\frac{\lambda_x}{\lambda_i}$ ratio. Consider the XPDC process $\omega_{X3} \to \omega_{X2} + \omega_{o1}$, which is physically equivalent to the Thomson scattering of X-rays by an atom illuminated with an optical field, Doppler-shifted sideband is induced by this process[23]. As elaborated in Section S4.1 of Supplementary Information, the XPDC cross section from nonlinear crystal scales as $\frac{d\sigma}{d\Omega}^{(2)} \propto 1/\omega_{o1}$ and especially favors conversion to low energy optical photons. Using the semiclassical treatment[15,23,24], we estimated the cross section of the instance in this work to be $\sim 1.9 \times 10^3$ fm$^2$ (see Section S4.1), which can be comparable to that of the Thomson scattering. We also show in Section S4.1 of Supplementary Information that the quasi-degenerate XPDC to pairs of two X-ray photons has cross section that is four orders of magnitude lower than the XPDC to pairs of X-ray and optical photons. Moreover the strong absorption of X-ray photons by the diamond crystal[8] and radiation damage caused deterioration of phase matching condition can significantly suppress the XPDC efficiency.

To overcome this problem, we could use radiation-hard multilayer photonic crystal to enhance quantum efficiency of XPDC, or to use beam of single electrons as nonlinear medium for XPDC through the three-color Kapitza-Dirac-like mechanism (Fig. 5)[25]. In Section S4.2 of Supplementary Information [15], we used perturbation theory[25–27] to find the probability of the desired XPDC process under the phase matching condition $\vec{k}_{X3} = \vec{k}_{o1} + \vec{k}_{X2}$ as

$$P(\omega_{o1}, \omega_{X2}, \omega_{X3}) = \left| \frac{v_z E_{o1} E_{X2} E_{X3}}{2c^2 \omega_{o1} \omega_{X2} \omega_{X3}} \right|^2 \delta(E_{fi}). \tag{12}$$

In the three-color Kapitza-Dirac-like process, the electrons are scattered $|\vec{p}_i\rangle \to |\vec{p}_f\rangle = |\vec{p}_i \pm (\vec{k}_{o1} + \vec{k}_{X2} + \vec{k}_{X3})\rangle$, and $\delta(E_{fi}) = \delta(p_f^2/2 - p_i^2/2)$ ensures the energy conservation of the electrons, and $v_z$ is the initial velocity of the electrons along the $z$ axis. With electric field intensity $I \sim 10^{18}$ W/cm$^2$ well below the QED critical intensity and moderate electron velocity in non-relativistic regime, the XPDC probability could reach $10^{-5}$.

In conclusion, we have theoretically described a mechanism to realize X-ray diffraction with an atomic resolution by two-color entanlged ghost diffraction. Because the sample is irradiated by photons of optical wavelength, the proposed scheme can be free of radiation damage by X-ray photons. In principle the proposed scheme could also be applied for single molecules to determine the molecular structure using phase retrieval techniques of coherent diffractive imaging. Moreover, achieving resolution on a much smaller length scale than the wavelength of the illuminating photons using quantum product state of fundamental particle entanglement opens the possibility for future development of quantum optics based imaging techniques, such as using entangled photon-electron, electron-electron pair or electron-anti-neutrino pair from $\beta$-decay.

## Acknowledgements

We thank the Hamburg Centre for Ultrafast Imaging for financial support. We thank Todd Martínez, Jochen Schneider, Nina Rohringer, Ivan Vartanyants, Ralf Röhlsberger, Robin Santra, Oriol Vendrell, Jerome Hastings, David Reis, Lida Zhang, Xiaolei Zhu, Tao Peng, Kareem Hegazy, Matthew Ware, James Cryan and Andreas Kaldun for commenting on various aspects of the study. Z.L. is thankful to the Volkswagen Foundation for a Paul-Ewald postdoctoral fellowship. Partial financial support from the Czech Ministry of Education (Grants LB15013 and LM2015083) is acknowledged by N.M..

## Author contributions statement

Z.L. conceived the method, Z.L. and N.M. conducted the theoretical investigation. All authors analyzed the results and reviewed the manuscript.

## Additional information

The authors declare no competing financial interest.

## References

1. Barty, A. *et al.* Self-terminating diffraction gates femtosecond X-ray nanocrystallography measurements. *Nature Photonics* **6**, 35 (2012).
2. Spence, J. X-ray imaging: Ultrafast diffract-and-destroy movies. *Nature Photonics* **2**, 390 (2008).
3. Nishino, Y., Takahashi, Y., Imamoto, N., Ishikawa, T. & Maeshima, K. Three-dimensional visualization of a human chromosome using coherent X-Ray diffraction. *Phys. Rev. Lett.* **102**, 018101 (2009).
4. Quiney, H. & Nugent, K. A. Biomolecular imaging and electronic damage using X-ray free-electron lasers. *Nature Phys.* **7**, 142 (2011).
5. Fratalocchi, A. & Ruocco, G. Single-molecule imaging with x-ray free-electron lasers: Dream or reality? *Phys. Rev. Lett.* **106**, 105504 (2011).
6. Glover, T. E. *et al.* X-ray and optical wave mixing. *Nature* **488**, 603 (2012).
7. Tamasaku, K., Sawada, K., Nishibori, E. & Ishikawa, T. Visualizeing the local optical response to extreme-ultraviolet radiation. *Nature Phys.* **7**, 705 (2011).
8. Shwartz, S. *et al.* X-ray parametric down-conversion in the langevin regime. *Phys. Rev. Lett.* **109**, 013602 (2012).
9. Pittman, T. B., Shih, Y. H., Strekalov, D. V. & Sergienko, A. V. Optical imaging by means of two-photon quantum entanglement. *Phys. Rev. A* **52**, R3429 (1995).
10. Strekalov, A., Sergienko, A. V., Klyshko, D. N. & Shih, Y. H. Observation of two-photon "ghost" interference and diffraction. *Phys. Rev. Lett.* **74**, 3600 (1995).
11. Scarcelli, G., Berardi, V. & Shih, Y. H. Can two-photon correlation of chaotic light be considered as correlation of intensity fluctuations? *Phys. Rev. Lett.* **96**, 063602 (2006).
12. Rubin, M. H. & Shih, Y. H. Resolution of ghost imaging for nondegenerate spontaneous parameteric down-conversion. *Phys. Rev. A* **78**, 033836 (2008).
13. Yu, H. *et al.* Fourier-transform ghost imaging with hard X rays. *Phys. Rev. Lett.* **117**, 113901 (2016).

14. Pelliccia, D., Rack, A., Scheel, M., Cantelli, V. & Paganin, D. M. Experimental x-ray ghost imaging. *Phys. Rev. Lett.* **117**, 113902 (2016).
15. *Supplementary Information* .
16. Shwartz, S. & Bömer, C. *Private communication* .
17. Tamasaku, K. & Ishikawa, T. Interference between compton scattering and x-ray parametric down-conversion. *Phys. Rev. Lett.* **98**, 244801 (2007).
18. Als-Nielsen, J. & McMorrow, D. *Elements of modern X-ray physics* (John Wiley & Sons, 2011).
19. Einstein, A., Podolsky, B. & Rosen, N. Can quantum-mechanical description of physical reality be considered complete? *Phys. Rev.* **47**, 777 (1935).
20. Klyshko, D. N. A simple method of preparing pure states of an optical field, of implementing the einstein-podolsky-rosen experiment, and of demonstrating the complementarity principle. *Usp. Fiz. Nauk* **154**, 133 (1988).
21. Van Hove, L. Correlations in space and time and born approximation scattering in systems of interacting particles. *Phys. Rev.* **95**, 249 (1954).
22. Adams, B. W. *et al.* X-ray quantum optics. *J. Mod. Opt.* **60**, 2 (2013).
23. Freund, I. & Levine, B. F. Optically modulated X-Ray diffraction. *Phys. Rev. Lett.* **25**, 1241–1245 (1970).
24. Zambianchi, P. Nonlinear response functions for X-Ray laser pulses. In *Nonlinear Optics, Quantum Optics, and Ultrafast Phenomena with X-Rays*, 287–302 (Springer Verlag, 2003).
25. Batelaan, H. *Colloquium*: Illuminating the Kapitza-Dirac effect with electron matter optics. *Rev. Mod. Phys.* **79**, 929–941 (2007).
26. Freimund, D. L., Aflatooni, K. & Batelaan, H. Observation of the Kapitza-Dirac effect. *Nature* **413**, 142–143 (2001).
27. Smirnova, O., Freimund, D. L., Batelaan, H. & Ivanov, M. Kapitza-Dirac diffraction without standing waves: Diffraction without a grating? *Phys. Rev. Lett.* **92**, 223601 (2004).

# SUPPLEMENTARY INFORMATION
# Radiation damage free two-color X-ray ghost diffraction with atomic resolution

**Zheng Li**[1,2,3,*], **Nikita Medvedev**[1,4], **Henry N. Chapman**[1,2,5], **and Yanhua Shih**[6,*]

[*]zheng.li@desy.de
[1]Center for Free-Electron Laser Science, DESY, Notkestraße 85, D-22607 Hamburg, Germany
[2]Hamburg Centre for Ultrafast Imaging, Luruper Chaussee 149, D-22761 Hamburg, Germany
[3]SLAC National Accelerator Laboratory, Menlo Park, California 94025, USA
[4]Institute of Physics and Institute of Plasma Physics, Academy of Science of Czech Republic, Na Slovance 1999/2, 18221 Prague 8, Czech Republic
[5]Department of Physics, University of Hamburg, Jungiusstraße 9, D-20355 Hamburg, Germany
[*]shih@umbc.edu
[6]Department of Physics, University of Maryland, Baltimore County, Baltimore, Maryland 21250, USA

## ABSTRACT

The supplementary information is divided into the following sections. Section S1 describes explicitly the analogous Bragg equation (Eq. 1 of the main text). Section S2 illustrate a simple kinematic description of Laue diffraction for the completeness of the theory presented in Section S1. Section S3 treats the broadening effect of Bragg peaks. Section S4 elaborates the estimation of XPDC efficiency for the two proposed schemes, the one using crystal and the another using single electron as the nonlinear medium. Section S5 explains the connection of the Thomson and the Rayleigh scattering in the X-ray and optical regime.

**Table S1.** Symbol table.

| Symbol | Explanation |
|---|---|
| $R_c$ | counting rate of joint photon detection |
| $|\Psi\rangle$ | two-photon state |
| $\hat{\rho}$ | density matrix of the two-photon state on the output plane of the nonlinear medium |
| $\hat{E}_A^{(+)}$ | positive frequency part of the quantized photon field on the plane of bucket detector A |
| $\hat{E}_B^{(+)}$ | positive frequency part of the quantized photon field on the plane of pixel detector B |
| $\hat{E}_A^{(-)}$ | negative frequency part of the quantized photon field on the plane of bucket detector A |
| $\hat{E}_B^{(-)}$ | negative frequency part of the quantized photon field on the plane of pixel detector B |
| $\hat{a}_p^\dagger(\vec{k})$ | photon creation operator for specific mode $\vec{k}$ and channel $p$, $p = s, i$ |
| $\hat{a}_p(\vec{k})$ | photon annihilation operator for specific mode $\vec{k}$ and channel $p$, $p = s, i$ |
| $\vec{k}_s$ | wave vector of the signal photon |
| $\vec{k}_i$ | wave vector of the idler photon |
| $k_s$ | $\left|\vec{k}_s\right|$ |
| $k_i$ | $\left|\vec{k}_i\right|$ |
| $\vec{\kappa}_s$ | transverse momentum of the signal photon |
| $\vec{\kappa}_i$ | transverse momentum of the idler photon |
| $\vec{Q}$ | momentum transfer of the photon-atom scattering |
| $\lambda_s$ | wavelength of the signal photon |
| $\lambda_i$ | wavelength of the idler photon |
| $\vec{\rho}_A$ | position vector on the plane of bucket detector A |
| $\vec{\rho}_B$ | position vector on the plane of pixel detector B |
| $\vec{\rho}_a$ | position vector on the crystal plane of specific Miller index |
| $\vec{\rho}_s$ | position vector for the signal photon on the output plane of the nonlinear medium |
| $\vec{\rho}_{s'}$ | position vector for the idler photon on the output plane of the nonlinear medium |
| $\vec{d}$ | distance vector between two atoms in the crystal plane of specific Miller index |
| $d$ | $\left|\vec{d}\right|$ |
| $d_s$ | distance from the output plane of the nonlinear medium to the lattice plane of the crystal |
| $L_s$ | distance from the lattice plane to the bucket detector A |
| $D_i$ | distance from the output plane of the nonlinear medium to the pixel detector B |
| $R_s$ | $d_s + \frac{\lambda_i}{\lambda_s} D_i$ |
| $z_s$ | $d_s + L_s$ |
| $z_i$ | $D_i$ |
| $\omega_s$ | frequency of the signal photon |
| $\omega_i$ | frequency of the idler photon |
| $\Omega_s$ | central frequency of the signal photon subject to phase matching condition |
| $\Omega_i$ | central frequency of the idler photon subject to phase matching condition |
| $U_{AB}$ | joint photon detection amplitude |
| $V(\vec{\rho}_A, \vec{\rho}_B)$ | interference kernel in the joint photon detection amplitude |
| $G(|\vec{\alpha}|, \beta)$ | $e^{i\frac{\beta}{2}|\vec{\alpha}|^2}$ |
| $\delta$ | difference of optical path lengths |
| $\theta$ | reflection angle of the optical photon from the crystal |
| $\psi_0$ | photon-atom scattering amplitude |
| $\tilde{m}$ | magnification factor |
| $\mathscr{L}(\theta)$ | line shape of Bragg peak with reflection angle $\theta$ |
| $L_c$ | length of a nanocrystal |

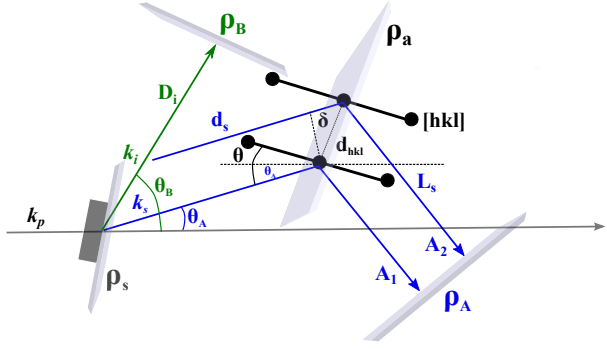

**Figure S1.** Proposed layout for two-color two-photon ghost diffraction of X-ray and optical photon pair. $\vec{\rho}_A$, $\vec{\rho}_B$, $\vec{\rho}_s$, $\vec{\rho}_a$ are vectors on the planes of the bucket detector A, the pixel detector B, the X-ray PDC source and atoms in the sample. The path of the optical photon in the arm A is labeled by blue lines, with the distance $d_s$ from the XPDC plane to the crystal plane and the distance from $L_s$ to the bucket detector A. The path of the X-ray photon in the arm B is labeled by green line, with the distance $D_i$ from the XPDC plane to the pixel detector B. $k_p$, $k_s$ and $k_i$ are the momentum vectors of the pump, the signal (optical) and the idler (X-ray) photons.

## S1  Bragg equation for two-color two-photon diffraction

We assume paraxial approximation for the wave vectors of modes $|\vec{k}\rangle$ of the photon field, $\vec{k} = \sqrt{k^2 - \kappa^2}\hat{e}_z + \vec{\kappa}$, with $k = \frac{\omega}{c} \gg \kappa$ and $\vec{\kappa} = (k_x, k_y, 0)$. The two-photon coincidence counting rate of the bucket detector A and the pixel detector can be written as[1,2,3],

$$R_c(\vec{\rho}_B) = \frac{1}{T}\int dt_A dt_B S(t_B, t_A)\sigma_B \mathrm{tr}_{\vec{\rho}_A}\left[\hat{E}_A^{(-)}\hat{E}_B^{(-)}\hat{E}_B^{(+)}\hat{E}_A^{(+)}\hat{\rho}\right], \tag{S1}$$

where $S(t_B, t_A)$ is the coincidence time function that vanishes unless $0 \leq t_B - t_A \leq T$ and can be approximated as a rectangular function, $\vec{\rho}_A$ and $\vec{\rho}_B$ are vectors on the plane of the bucket detector A with an area $\sigma_A$ and the plane of the pixel detector B with an area $\sigma_B$ respectively. $\hat{\rho}$ is the density matrix of the two-photon state on the output plane of the nonlinear crystal. $\hat{E}_j^{(+)}$, $j = A, B$ is the positive frequency part of the photon field, and $\hat{E}_j^{(-)} = (\hat{E}_j^{(+)})^\dagger$. And $\mathrm{tr}_{\vec{\rho}_A}[\cdots]$ denotes the trace and coherent summation over $\vec{\rho}_A$. We compute the photon fields at the plane of the bucket detector A and the pixel detector B, $\hat{E}_A^{(+)}$ and $\hat{E}_B^{(+)}$ as,

$$\begin{aligned}
\hat{E}_A^{(+)}(\vec{\kappa}_s, \vec{k}_s, \vec{\rho}_A, z_s, \omega_s, t_A) &= \sum_{\vec{k}_s} g(\vec{\kappa}_s, \omega_s, \vec{\rho}_A, z_s)\hat{a}_s(\vec{k}_s)e^{-i\omega_s t_A} \\
\hat{E}_B^{(+)}(\vec{\kappa}_i, \vec{k}_i, \vec{\rho}_B, z_i, \omega_i, t_B) &= \sum_{\vec{k}_i} g(\vec{\kappa}_i, \omega_i, \vec{\rho}_B, z_i)\hat{a}_i(\vec{k}_i)e^{-i\omega_i t_B},
\end{aligned} \tag{S2}$$

where $z_s = d_s + L_s$ and $z_i = D_i$ are the full optical path lengths for the signal and the idler photon. $\hat{a}_p^\dagger(\vec{k})$ and $\hat{a}_p(\vec{k})$ are operators of signal and idler photon fields in a specific mode at the output plane of the nonlinear crystal, with the commutator relation,

$$\left[\hat{a}_p(\vec{k}), \hat{a}_q^\dagger(\vec{k}')\right] = \delta_{p,q}\delta_{\omega,\omega'}\delta_{\vec{\kappa},\vec{\kappa}'}. \tag{S3}$$

$g(\vec{\kappa}, \omega, \vec{\rho}, z)$ is the Green's function for a specific mode of photon field.

Assume two atoms in two lattice planes of Miller index $[hkl]$ with the distance $d = d_{hkl}$ are in a plane $a$, $\vec{\rho}_a$ is a vector in this plane (Fig. S1), and the photon-atom scattering amplitude is $t(\vec{\rho}_a)$, we shows that $\hat{E}_A^{(+)}$ and $\hat{E}_B^{(+)}$ can be written as[2],

$$\begin{aligned}
\hat{E}_A^{(+)} &\equiv \hat{E}_A^{(+)}(\vec{\kappa}_s, \vec{k}_s, \vec{\rho}_A, z_s, \omega_s, t_A) = \sum_{\vec{k}_s}\int d^2\vec{\rho}_{s'}\int d^2\vec{\rho}_a \frac{-ik_s}{2\pi d_s}e^{ik_s d_s}e^{i\vec{\kappa}_s\cdot\vec{\rho}_{s'}}e^{i\frac{k_s}{2d_s}|\vec{\rho}_a - \vec{\rho}_{s'}|^2}t(\vec{\rho}_a) \\
&\quad \times \frac{-ik_s}{2\pi L_s}e^{ik_s L_s}e^{i\frac{k_s}{2L_s}|\vec{\rho}_A - \vec{\rho}_a|^2}e^{-i\omega_s t_A}\hat{a}_s(\vec{k}_s) \\
\hat{E}_B^{(+)} &\equiv \hat{E}_B^{(+)}(\vec{\kappa}_i, \vec{k}_i, \vec{\rho}_B, z_i, \omega_i, t_B) = \sum_{\vec{k}_i}\int d^2\vec{\rho}_s \frac{-ik_i}{2\pi D_i}e^{ik_i D_i}e^{i\vec{\kappa}_i\cdot\vec{\rho}_s}e^{i\frac{k_i}{2D_i}|\vec{\rho}_s - \vec{\rho}_B|^2}e^{-i\omega_i t_B}\hat{a}_i(\vec{k}_i).
\end{aligned} \tag{S4}$$

Denote $G(|\vec{\alpha}|,\beta) = e^{i\frac{\beta}{2}|\vec{\alpha}|^2}$, its two-dimensional Fourier transformation is

$$\int d^2\vec{\alpha}\, G(|\vec{\alpha}|,\beta) e^{i\vec{\gamma}\cdot\vec{\alpha}} = i\frac{2\pi}{\beta} G(|\vec{\gamma}|, -\frac{1}{\beta}). \tag{S5}$$

The photon fields $\hat{E}_A^{(+)}$ and $\hat{E}_B^{(+)}$ can be then written as

$$\begin{aligned}
\hat{E}_A^{(+)} &= \sum_{\vec{k}_s} \int d^2\vec{\rho}_a \frac{1}{i\lambda_s L_s} t(\vec{\rho}_a) G(|\vec{\rho}_a - \vec{\rho}_A|, \frac{k_s}{L_s}) e^{i\vec{\kappa}_s\cdot\vec{\rho}_a} G(|\vec{\kappa}_s|, -\frac{d_s}{k_s}) e^{i(k_s z_s - \omega_s t_A)} \hat{a}_s(\vec{k}_s) \\
\hat{E}_B^{(+)} &= \sum_{\vec{k}_i} G(|\vec{\kappa}_i|, -\frac{D_i}{k_i}) e^{i\vec{\kappa}_i\cdot\vec{\rho}_B} e^{i(k_i z_i - \omega_i t_B)} \hat{a}_i(\vec{k}_i).
\end{aligned} \tag{S6}$$

With Eq. S6, we can obtain explicitly the second-order coherence function in Eq. S1,

$$\begin{aligned}
G_{AB} &= \text{tr}_{\vec{\rho}_A}\left[\hat{E}_A^{(-)} \hat{E}_B^{(-)} \hat{E}_B^{(+)} \hat{E}_A^{(+)} \hat{\rho}\right] \\
&= \left|\int d^2\vec{\rho}_A \langle 0|\hat{E}_B^{(+)} \hat{E}_A^{(+)}|\Psi\rangle\right|^2 \\
&\equiv \left|\int d^2\vec{\rho}_A U_{AB}\right|^2,
\end{aligned} \tag{S7}$$

where $|\Psi\rangle$ is the two-photon state vector, and $U_{AB}$ is the joint photon detection amplitude. Without loss of generality, we ignore here the reciprocal lattice vector $\vec{G}$ of the nonlinear crystal, which can be added for a given experimental configuration. Suppose the phase matching condition is satisfied for $\omega_p n_p(\omega_p) - \Omega_s n_s(\Omega_s) - \Omega_i n_i(\Omega_i) = 0$, with $K_j = \frac{\Omega_j}{c}$, and deviation from the central frequency $\omega_j = \Omega_j + \nu_j$, $j = s,i$. Frequency and spatial filtering guarantees $\nu_j \ll \Omega_j$ and $\kappa_j \ll K_j$, thus we have

$$\vec{k}_j = \vec{\kappa}_j + (K_j + \frac{\nu_j}{c} + \frac{\nu_j^2}{2\Omega_j c} - \frac{\kappa_j^2}{2K_j})\hat{e}_z \simeq \vec{\kappa}_j + (K_j + \frac{\nu_j}{c})\hat{e}_z. \tag{S8}$$

Taken Eqs. S6 and S7, we have

$$\begin{aligned}
U_{AB} &= \frac{1}{i\lambda_s L_s} e^{i(K_s z_s + K_i z_i - \Omega_s t_A - \Omega_i t_B)} \left[\sum_{\vec{k}_s,\vec{k}_i} g'_A(\vec{\rho}_A, \nu_s, \vec{\kappa}_s) g'_B(\vec{\rho}_B, \nu_i, \vec{\kappa}_i) \langle 0|\hat{a}_s(\vec{k}_s)\hat{a}_i(\vec{k}_i)|\Psi\rangle\right] \\
&\equiv \frac{1}{i\lambda_s L_s} e^{i(K_i z_i - \Omega_s t_A - \Omega_i t_B)} U'.
\end{aligned} \tag{S9}$$

Using a first-order perturbation theory, the two-photon amplitude from a PDC process is shown to be[2,4],

$$\langle 0|\hat{a}_s(\vec{k}_s)\hat{a}_i(\vec{k}_i)|\Psi\rangle = -i(2\pi)^3 \gamma \delta(\nu_s + \nu_i) \delta(\vec{\kappa}_s + \vec{\kappa}_i) \text{sinc}(\nu_s D_{si}\frac{L}{2}), \tag{S10}$$

where $\gamma = \frac{\chi^{(2)} E_p L}{2 U_s U_i} \sqrt{\frac{\Omega_s \Omega_i T_s T_i c^2}{n_s n_i}}$, $D_{si} = \frac{1}{U_s} - \frac{1}{U_i}$, $U_s$ and $U_i$ are the group velocity of the signal and the idler photons inside the nonlinear crystal of length $L$. $\chi^{(2)}$ is the second-order susceptibility of the nonlinear crystal, $U_j$ is the group velocity of the signal and the idler photons inside the nonlinear crystal of length $L$, $T_j$ are their transmission coefficients, and $E_p$ is the strength of the pump field. Using the relation

$$\sum_{\vec{k}_s} \to \frac{V_Q}{(2\pi)^3} \int d\vec{k}_s = \frac{V_Q}{(2\pi)^3} \int d^2\vec{\kappa}_s \int d\omega_s \frac{dk_{s,z}}{d\omega_s} = \frac{V_Q}{(2\pi)^3 c} \int d^2\vec{\kappa}_s \int d\nu_s \tag{S11}$$

$$\hat{a}_s(\vec{k}_s) \to \frac{c}{V_Q} \hat{a}_s(\vec{\kappa}_s, \nu_s), \tag{S12}$$

where $V_Q$ is the quantization volume. We can write $U'$ in Eq. S9 as

$$\begin{aligned}
U' &= -i(2\pi)^{-3} \gamma \int d\nu_s e^{i\nu_s \tau_{BA}} \text{sinc}(\nu_s D_{si}\frac{L}{2}) \int d^2\vec{\rho}_a \int d^2\vec{\kappa}_s G(|\vec{\kappa}_s|, -\frac{D_i}{k_i}) e^{-i\vec{\kappa}_s\cdot\vec{\rho}_B} G(|\vec{\kappa}_s|, -\frac{d_s}{k_s}) e^{i\vec{\kappa}_s\cdot\vec{\rho}_a} \\
&\quad \times t(\vec{\rho}_a) G(|\vec{\rho}_a|, \frac{k_s}{L_s}) G(|\vec{\rho}_A|, \frac{k_s}{L_s}) e^{-i\frac{k_s}{L_s}\vec{\rho}_a\cdot\vec{\rho}_A} e^{iK_s z_s} \\
&\equiv -i\gamma \int d\nu_s e^{i\nu_s \tau_{BA}} \text{sinc}(\nu_s D_{si}\frac{L}{2}) \int d^2\vec{\rho}_a I_{\vec{\kappa}_s} t(\vec{\rho}_a) G(|\vec{\rho}_a|, \frac{k_s}{L_s}) G(|\vec{\rho}_A|, \frac{k_s}{L_s}) e^{-i\frac{k_s}{L_s}\vec{\rho}_a\cdot\vec{\rho}_A} e^{iK_s z_s},
\end{aligned} \tag{S13}$$

where $\tau_{BA} = \tau_B - \tau_A$, $\tau_i = t_i - \frac{r_i}{c}$, $i = A, B$, and $r_i$ are the full lengths of the optical paths. Define the effective sample-to-pixel detector path length $R_s = d_s + \frac{\lambda_i}{\lambda_s} D_i$, the integral $I_{\vec{\kappa}_s}$ over transverse momentum $\vec{\kappa}_s$ in Eq. S13 can be rewritten as

$$
\begin{aligned}
I_{\vec{\kappa}_s} &= \frac{1}{(2\pi)^3} \int d^2 \vec{\kappa}_s \, G(|\vec{\kappa}_s|, -(\frac{D_i}{k_i} + \frac{d_s}{k_s})) e^{i\vec{\kappa}_s \cdot (\vec{\rho}_a - \vec{\rho}_B)} \\
&= \frac{1}{(2\pi)^3} (-i) \frac{2\pi}{\frac{D_i}{k_i} + \frac{d_s}{k_s}} G(|\vec{\rho}_a - \vec{\rho}_B|, \frac{1}{\frac{D_i}{k_i} + \frac{d_s}{k_s}}) \\
&= \frac{1}{2\pi} \frac{1}{i\lambda_s R_s} G(|\vec{\rho}_a - \vec{\rho}_B|, \frac{K_s}{R_s}),
\end{aligned}
\qquad (S14)
$$

because $\frac{D_i}{k_i} + \frac{d_s}{k_s} = \frac{1}{2\pi}(D_i \lambda_i + d_s \lambda_s) = \frac{\lambda_s}{2\pi}(D_i \frac{\lambda_i}{\lambda_s} + d_s) = \frac{R_s}{K_s}$. The amplitude for joint photon detection at the bucket detector A and the pixel detector B is then

$$
\begin{aligned}
U_{AB} &= \langle 0|\hat{E}_B^{(+)} \hat{E}_A^{(+)}|\Psi\rangle = -\frac{\gamma}{\lambda_s L_s} e^{i(K_i z_i - \Omega_s t_A - \Omega_i t_B)} \int d\nu_s e^{i\nu_s \tau_{BA}} \mathrm{sinc}(\nu_s D_{si} \frac{L}{2}) \\
&\quad \times \left\{ \frac{e^{iK_s z_s}}{2\pi i \lambda_s R_s} G(|\vec{\rho}_A|, \frac{K_s}{L_s}) G(|\vec{\rho}_B|, \frac{K_s}{R_s}) \int d^2 \vec{\rho}_a t(\vec{\rho}_a) G(|\vec{\rho}_a|, K_s(\frac{1}{L_s} + \frac{1}{R_s})) e^{-iK_s(\frac{\vec{\rho}_B}{R_s} + \frac{\vec{\rho}_A}{L_s}) \cdot \vec{\rho}_a} \right\} \\
&\equiv -\frac{\gamma}{\lambda_s L_s} e^{i(K_i z_i - \Omega_s t_A - \Omega_i t_B)} \int d\nu_s e^{i\nu_s \tau_{BA}} \mathrm{sinc}(\nu_s D_{si} \frac{L}{2}) V(\vec{\rho}_A, \vec{\rho}_B),
\end{aligned}
\qquad (S15)
$$

where $V(\vec{\rho}_A, \vec{\rho}_B)$ is the interference kernel of the joint photon detection amplitude $U_{AB}$ that characterizes the formation of the two-photon diffraction pattern. We now consider the scattering of the optical photon with the atoms in a lattice by integrating over $\vec{\rho}_a$, and the collection of optical photons at the bucket detector A by integrating over $\vec{\rho}_A$ in the detector plane, because the bucket detector collects photons without distinguishing their actual position. We expect that diffraction patterns can be formed on the image plane of the pixel detector B, with intensity $I(\vec{\rho}_B)$ and joint photon counting rate $R_c(\vec{\rho}_B)$ at $\vec{\rho}_B$

$$
I(\vec{\rho}_B) = R_c(\vec{\rho}_B) T \sim \left| \int d^2 \vec{\rho}_A V(\vec{\rho}_A, \vec{\rho}_B) \right|^2.
\qquad (S16)
$$

To obtain the analogous Bragg equation for two-color two-photon ghost diffraction, we consider two optical paths $A_1$ and $A_2$ of the optical photons that scatter with two atoms in two lattice planes with distance $d \equiv d_{hkl}$ (Fig. S1). For optical path $A_1$, we have

$$
\begin{aligned}
L_s^{(1)} &= L_s \\
R_s^{(1)} &= d_s + \frac{\lambda_i}{\lambda_s} D_i \equiv R_s \\
z_s^{(1)} &= d_s + D_i \equiv z_s,
\end{aligned}
\qquad (S17)
$$

and for optical path $A_2$

$$
\begin{aligned}
L_s^{(2)} &= L_s \\
R_s^{(2)} &= (d_s + 2\delta) + \frac{\lambda_i}{\lambda_s} D_i \equiv R_s + 2\delta \\
z_s^{(2)} &= (d_s + 2\delta) + D_i \equiv z_s + 2\delta,
\end{aligned}
\qquad (S18)
$$

where $\delta = d \sin\theta$. Thus the effective two-photon joint detection amplitude consists of contributions from the two optical paths, $V(\vec{\rho}_A, \vec{\rho}_B) = V^{(1)}(\vec{\rho}_A, \vec{\rho}_B) + V^{(2)}(\vec{\rho}_A, \vec{\rho}_B)$, and $t(\vec{\rho}_a) = t^{(1)}(\vec{\rho}_a) + t^{(2)}(\vec{\rho}_a) = \psi_0[\delta(\vec{\rho}_a) + \delta(\vec{\rho}_a - \vec{d})]$, where $\vec{d} = (d, 0)$. We write $V^{(1)}(\vec{\rho}_A, \vec{\rho}_B)$ and $V^{(2)}(\vec{\rho}_A, \vec{\rho}_B)$ explicitly as

$$
\begin{aligned}
V^{(1)}(\vec{\rho}_A, \vec{\rho}_B) &= \frac{\psi_0 e^{iK_s z_s}}{2\pi i \lambda_s R_s} G(|\vec{\rho}_A|, \frac{K_s}{L_s}) G(|\vec{\rho}_B|, \frac{K_s}{R_s}) \\
V^{(2)}(\vec{\rho}_A, \vec{\rho}_B) &= \frac{\psi_0 e^{iK_s(z_s + 2\delta)}}{2\pi i \lambda_s (R_s + 2\delta)} G(|\vec{\rho}_A|, \frac{K_s}{L_s}) G(|\vec{\rho}_B|, \frac{K_s}{R_s + 2\delta}) \\
&\quad \times \left\{ G\left(|\vec{d}|, K_s(\frac{1}{L_s} + \frac{1}{R_s + 2\delta})\right) e^{-iK_s(\frac{\vec{\rho}_B}{R_s + 2\delta} + \frac{\vec{\rho}_A}{L_s}) \cdot \vec{d}} \right\}.
\end{aligned}
\qquad (S19)
$$

Define $W^{(p)}(\vec{\rho}_B) = \int d^2\vec{\rho}_A V^{(p)}(\vec{\rho}_A, \vec{\rho}_B)$, we have

$$
\begin{aligned}
W^{(1)}(\vec{\rho}_B) &= \frac{\psi_0 L_s}{2\pi R_s} e^{iK_s z_s} G(|\vec{\rho}_B|, \frac{K_s}{R_s}) \\
W^{(2)}(\vec{\rho}_B) &= \frac{\psi_0 e^{iK_s(z_s+2\delta)}}{2\pi i \lambda_s (R_s + 2\delta)} G(|\vec{\rho}_B|, \frac{K_s}{R_s}) e^{-i\frac{K_s \delta}{R_s^2}|\vec{\rho}_B|^2} e^{i\frac{K_s}{2}(\frac{1}{L_s}+\frac{1}{R_s})|\vec{d}|^2} e^{-i\frac{K_s \delta}{R_s^2}|\vec{d}|^2} \\
&\quad \times e^{-iK_s \frac{\vec{\rho}_B}{R_s} \cdot \vec{d}} e^{i\frac{2K_s \delta}{R_s^2}\vec{\rho}_B \cdot \vec{d}} \int d^2\vec{\rho}_A G(|\vec{\rho}_A|, \frac{K_s}{L_s}) e^{-iK_s \frac{\vec{d}}{L_s} \cdot \vec{\rho}_A} \\
&= \frac{\psi_0 L_s}{2\pi R_s} e^{iK_s z_s} G(|\vec{\rho}_B|, \frac{K_s}{R_s}) e^{i\frac{K_s}{2R_s}(|\vec{d}|^2 - 2\vec{\rho}_B \cdot \vec{d})} e^{iK_s[2\delta(1 - \frac{|\vec{\rho}_B - \vec{d}|^2}{2R_s^2})]}.
\end{aligned}
\tag{S20}
$$

Take far-field approximation $\delta \ll R_s$ that is similar to the Laue diffraction, we have

$$
\begin{aligned}
|W(\vec{\rho}_B)|^2 &= \left| \int d^2\vec{\rho}_A V(\vec{\rho}_A, \vec{\rho}_B) \right|^2 \\
&= \left(\frac{\psi_0 L_s}{2\pi R_s}\right)^2 \left\{ 2 + e^{i\frac{K_s}{2R_s}(|\vec{d}|^2 - 2\vec{\rho}_B \cdot \vec{d})} e^{iK_s[2\delta(1 - \frac{|\vec{\rho}_B - \vec{d}|^2}{2R_s^2})]} + \text{c.c.} \right\}.
\end{aligned}
\tag{S21}
$$

Analogous to the procedure in Sec. S2, we obtain the Bragg equation for two-color two-photon ghost diffraction from Eq. S21 by requiring the $\delta$-dependent phase to vanish, i.e.

$$
\begin{aligned}
K_s \left[ 2\delta(1 - \frac{|\vec{\rho}_B - \vec{d}|^2}{2R_s^2}) \right] &= 2n\pi \\
\Rightarrow 2d \sin\theta \left[ 1 - \frac{|\vec{\rho}_B - \vec{d}|^2}{2\left(\frac{d_s}{D_i} + \frac{\lambda_i}{\lambda_s}\right)^2 D_i^2} \right] &= n\lambda_s ,
\end{aligned}
\tag{S22}
$$

with an integer $n$. Define the magnification factor $\tilde{m}$ as

$$
\tilde{m} \equiv \tilde{m}(\vec{\rho}_B, \vec{d}, d_s, D_i, \lambda_i, \lambda_s) = 1 - \frac{|\vec{\rho}_B - \vec{d}|^2}{2\left(\frac{d_s}{D_i} + \frac{\lambda_i}{\lambda_s}\right)^2 D_i^2},
\tag{S23}
$$

the Bragg equation can be then written as

$$
2\tilde{m} d \sin\theta = n\lambda_s .
\tag{S24}
$$

Provided the Bragg condition is satisfied, and define $\vec{\rho}_B = (x, 0)$ and $\vec{d} = (d, 0)$ for simplicity, we find

$$
\begin{aligned}
|W(\vec{\rho}_B)|^2 &= \left| \int d^2\vec{\rho}_A V(\vec{\rho}_A, \vec{\rho}_B) \right|^2 \\
&= \left(\frac{\psi_0 L_s}{\pi R_s}\right)^2 \cos^2\left[\frac{2\pi}{\lambda_s R_s} d(x - \frac{d}{2})\right]
\end{aligned}
\tag{S25}
$$

which forms a broad and flat background. Using Eqs. S1, S7, S15 and S25, we reach the final equation for the counting rate of joint photon detection

$$
\begin{aligned}
R_c(\vec{\rho}_B) &= \frac{1}{T} \int dt_A dt_B S(t_B, t_A) \frac{\sigma_B \gamma^2}{\lambda_s^2 L_s^2} \left| \int d\nu_s e^{i\nu_s \tau_{BA}} \text{sinc}(\nu_s D_{si} \frac{L}{2}) \right|^2 \left| \int d^2\vec{\rho}_A V(\vec{\rho}_A, \vec{\rho}_B) \right|^2 \\
&= \frac{1}{T} \int dt_A dt_B S(t_B, t_A) \sigma_B \left(\frac{\gamma \psi_0}{\pi \lambda_s R_s}\right)^2 \left| \int d\nu_s e^{i\nu_s \tau_{BA}} \text{sinc}(\nu_s D_{si} \frac{L}{2}) \right|^2 \cos^2\left[\frac{2\pi}{\lambda_s R_s} d(x - \frac{d}{2})\right].
\end{aligned}
\tag{S26}
$$

Eq. S26 can be simplified using the relation

$$
\begin{aligned}
&\int_{-\infty}^{\infty} d\nu_s e^{i\nu_s \tau_{BA}} \text{sinc}(\nu_s D_{si} \frac{L}{2}) \\
&= \frac{2\pi}{D_{si}L} \text{rect}(\frac{2\tau_{AB}}{D_{si}L}) \\
&= \begin{cases} \frac{2\pi}{D_{si}L}, & -\frac{D_{si}L}{2} \leq \tau_{AB} \leq \frac{D_{si}L}{2} \\ 0, & |\tau_{AB}| > \frac{D_{si}L}{2} \end{cases},
\end{aligned}
\tag{S27}
$$

and the counting rate of joint-detection can be written as

$$R_c(\vec{\rho}_B) = \frac{1}{T}\int dt_A dt_B S(t_B,t_A)\sigma_B \left(\frac{2\gamma\psi_0}{D_{si}L\lambda_s R_s}\right)^2 \text{rect}(\frac{2\tau_{AB}}{D_{si}L})\cos^2\left[\frac{2\pi}{\lambda_s R_s}d(x-\frac{d}{2})\right]. \tag{S28}$$

## S2 Kinematic description of Laue diffraction

In Section S1, the Bragg condition for the two-color two-photon ghost diffraction is obtained from a kinematic scenario. Here we show the conventional Bragg equation can be obtained from similar procedure for the Laue diffraction with monochromatic X-ray beam.

According to Fourier optics, a light wave that passes the plane $z = 0$, and arrives at the plane $z = \Delta$ can be described by the Huygens principle as

$$\begin{aligned}
E(x,y,\Delta) &= \int dk_x dk_y e^{i(k_x x+k_y y)} e^{ik_z \Delta} \mathscr{F}[E(x,y,0)] \\
&= \int dk_x dk_y e^{i(k_x x+k_y y)} e^{i\Delta\sqrt{k^2-k_x^2-k_y^2}} \check{E}(k_x,k_y,0) \\
&\simeq \int dk_x dk_y e^{i(k_x x+k_y y)} e^{i\Delta\left(k-\frac{k_x^2+k_y^2}{2k}\right)} \check{E}(k_x,k_y,0) \\
&= \mathscr{F}^{-1}\left\{e^{i\Delta\left(k-\frac{k_x^2+k_y^2}{2k}\right)} \check{\psi}(k_x,k_y,0)\right\} \\
&= E(x,y,0) \otimes P(x,y,\Delta),
\end{aligned} \tag{S29}$$

where $P(x,y,\Delta)$ is

$$\begin{aligned}
P(x,y,\Delta) &= \mathscr{F}^{-1}\left\{e^{i\Delta k} e^{i\Delta\left(\frac{k_x^2+k_y^2}{2k}\right)}\right\} \\
&= -\frac{ike^{ik\Delta}}{2\pi\Delta}e^{i\frac{k(x^2+y^2)}{2\Delta}}.
\end{aligned} \tag{S30}$$

The physical scenario of Eq. S29 reflects a typical statement of Huygens principle, that (a) the Fourier transformation on the $z = 0$ plane makes a map to the momentum space $E(x,y,0) \to \check{E}(k_x,k_y,0)$, (b) each Fourier mode $\check{E}(k_x,k_y,0)$ corresponds to a sub-source that travels as a plane wave $e^{i(k_x x+k_y y)} e^{ik_z \Delta}$ to the $z = \Delta$ plane, and (c) an inverse Fourier transformation on the $z = \Delta$ plane gives the image $E(x,y,\Delta)$. For Laue diffraction, we follow a standard treatment by considering two optical paths 1 and 2 for photons that scatter off atoms in two lattice planes (Fig. S2(a)). Assume the photon scatters off the two atoms $O$ and $O'$ with amplitude $t^{(1)} = \psi_0\delta(x-d,y,0)$ and $t^{(2)} = \psi_0\delta(x,y,0)$, and denote $\delta = d\sin\theta$, we have

$$\begin{aligned}
E_1(x,y,\Delta) &= \psi_0 \otimes P(x-0,y-d,\Delta) = -\frac{ik\psi_0 e^{ik\Delta}}{2\pi\Delta}e^{i\frac{k}{2\Delta}[(x-d)^2+y^2]} \\
E_2(x,y,\Delta) &= \mathscr{F}^{-1}\left\{e^{ik_z\delta}\left[\int dx'dy' e^{-i(k_x x'+k_y y')}\psi_0\delta(x',y',0)\right] e^{ik(\Delta+\delta)} e^{-i(\Delta+\delta)\left(\frac{k_x^2+k_y^2}{2k}\right)}\right\} \\
&= -\frac{ik\psi_0 e^{ik(\Delta+2\delta)}}{2\pi(\Delta+2\delta)}e^{i\frac{k}{2(\Delta+2\delta)}(x^2+y^2)}.
\end{aligned} \tag{S31}$$

Thus the intensity of diffraction pattern $I(x)$ at the detector plane is given by

$$\begin{aligned}
I(x) &= |E(x,y,\Delta)|^2 = |E_1(x,y,\Delta)+E_2(x,y,\Delta)|^2 \\
&= \left(\frac{k\psi_0}{2\pi}\right)^2\left\{\frac{1}{\Delta^2} + \frac{1}{(\Delta+2\delta)^2} + \frac{1}{\Delta(\Delta+\delta)}\left[e^{i\frac{\pi}{\lambda}\left[\frac{(x-d)^2}{\Delta}-\frac{x^2}{\Delta+2\delta}\right]}e^{-i\frac{2\pi}{\lambda}2\delta} + \text{c.c.}\right]\right\} \\
&= \left(\frac{k\psi_0}{2\pi}\right)^2 I_1.
\end{aligned} \tag{S32}$$

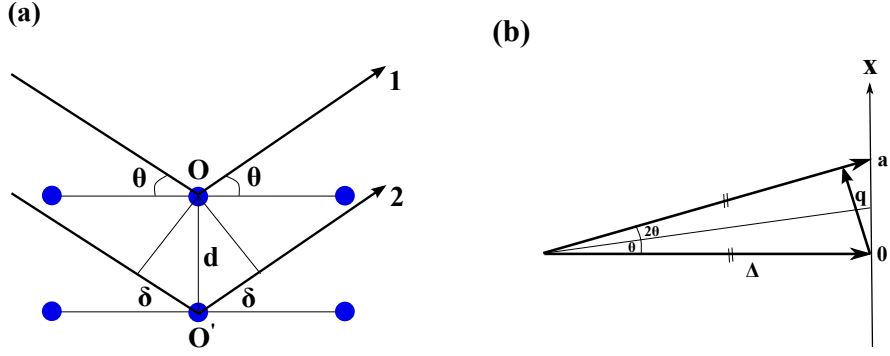

**Figure S2.** (a) Optical path diagram for Laue diffraction. $O$ and $O'$ are the positions of the atoms in adjacent crystal planes, $\theta$ is the reflection angle, and $\delta$ denotes the optical path length difference. (b) Momentum relation of Laue diffraction. $\Delta$ is the optical path length of the incident and scattered photons along their momentum vectors, and $\vec{q}$ is the momentum transfer.

For far-field diffraction $\Delta \gg 2\delta$, we have

$$
\begin{aligned}
I_1 &\simeq \frac{1}{\Delta^2}\left\{2 + 2\cos\left[\frac{\pi}{\lambda\Delta}[(x-d)^2 - x^2] - \frac{2\pi}{\lambda}2\delta\right]\right\} \\
&= \frac{2}{\Delta^2}\left\{1 + \cos\left[\frac{\pi}{\lambda\Delta}d(d-2x) - \frac{2\pi}{\lambda}2\delta\right]\right\} \\
&= \left(\frac{2}{\Delta}\right)^2 \cos^2\left[\frac{\pi}{\lambda\Delta}d(x-\frac{d}{2}) - \frac{\pi}{\lambda}2\delta\right],
\end{aligned}
\tag{S33}
$$

using the relation $1 + \cos\alpha = 2\cos^2\frac{\alpha}{2}$.

The Bragg condition is obtained by requiring the $\delta$-dependent phase in Eq. S33 to vanish,

$$
2\delta = 2d\sin\theta = n\lambda. \tag{S34}
$$

We obtain the diffraction pattern on the detector at the distance $\Delta$ from the sample,

$$
I(x) = \left(\frac{k\psi_0}{\pi\Delta}\right)^2 \cos^2\left[\frac{\pi}{\lambda\Delta}d(x-\frac{d}{2})\right], \tag{S35}
$$

with a period $a = \frac{2\lambda\Delta}{d}$. Meanwhile, we show that the kinematic description is consistent with the description of Laue diffraction in momentum space. The form factor $f(\vec{Q})$ is the Fourier transformation of the charge distribution. For simplicity, we model the atoms as point charges, thus

$$
f(\vec{Q}) = \int d\vec{r}\, e^{i\vec{Q}\cdot\vec{r}}\left[\delta(\vec{r}) + \delta(\vec{r}-\vec{d})\right] = 1 + e^{iQd}, \tag{S36}
$$

and the intensity of diffraction pattern is

$$
I \sim |f(\vec{Q})|^2 = 2(1+\cos Qd) = 4\cos^2\frac{Qd}{2}. \tag{S37}
$$

From Fig. S2(b), we can find that $\sin\theta \simeq \frac{x}{2\Delta}$ and

$$
Q \equiv |\vec{Q}| = 2|\vec{k}|\sin\theta \simeq 2\frac{2\pi}{\lambda}\frac{x}{2\Delta} = \frac{2\pi x}{\lambda\Delta} \tag{S38}
$$

Taken Eqs. S37 and S38, we obtain the intensity of diffraction pattern

$$I(x) \sim |f(\vec{Q})|^2 = 4\cos^2\left(\frac{\pi d}{\lambda \Delta}x\right),\tag{S39}$$

with a period $a = \frac{2\lambda\Delta}{d}$, which is consistent with Eq. S35 from the kinematic description of Laue diffraction.

## S3 Broadening of Bragg peaks

Analogous to the Scherrer equation, we calculate the width of the Bragg peaks of the two-color two-photon ghost diffraction, which determines the resolution of the proposed scheme. It is especially important for the application of XFEL to the diffraction of nanocrystals with a finite size. Supposing a nanocrystal of length $L_c$ that contains $N$ lattice planes with inter-plane distance $d$, such that $Nd = L_c$, the accumulated $\delta$-dependent phase for the diffraction from $N$ atoms can be calculated using Eq. S21 as

$$\begin{aligned}\Gamma(\theta) &= \sum_{p=0}^{N-1} e^{ipK_s\left[2\delta\left(1-\frac{|\vec{\rho}_B-\vec{d}|^2}{2R_s^2}\right)\right]} \\ &= Ne^{i\frac{N-1}{2}K_s(2\tilde{m}d\sin\theta)}\operatorname{sinc}\left[\frac{N}{2}K_s(2\tilde{m}d\sin\theta)\right].\end{aligned}\tag{S40}$$

Thus the line shape $\mathscr{L}(\theta)$ of a Bragg peak at a reflection angle $\theta$ is

$$\mathscr{L}(\theta) = \Gamma^2(\theta) = N^2\operatorname{sinc}^2\left[\frac{N}{2}K_s(2\tilde{m}d\sin\theta)\right].\tag{S41}$$

The width of the Bragg peak can be simply found through zeros of the line shape function as

$$\frac{N}{2}K_s(2\tilde{m}d\cos\theta)\Delta\theta = 2\pi.\tag{S42}$$

Defining the width of the Bragg peak as $B = 2\Delta\theta$, we find

$$B = \frac{2\lambda_s}{L_c \cos\theta \tilde{m}}.\tag{S43}$$

The factor $\frac{\lambda_s}{\tilde{m}}$ in the width of Bragg peaks guarantees that the two-color two-photon ghost diffraction will have a similar resolution to the Laue diffraction, i.e. on the atomic scale.

## S4 X-ray parametric down conversion through nonlinear crystal and single electron

In this section, we present details of the two proposed schemes for X-ray parametric down conversion (XPDC) using conventional nonlinear crystals or single electron as the nonlinear medium. We also present an estimation of the efficiency from experimental consideration for the XPDC of an X-ray photon of frequency $\omega_{X3}$ to a pair of X-ray and optical photons of frequencies $\omega_{X2}$ and $\omega_{o1}$. Both schemes rely dominantly on the figure eight motion of electrons turned by Lorentz force. For the nonlinear crystal based XPDC, the phase matching condition relies critically on the well-defined lattice vectors of reciprocal space, thus the structural damage to the crystal inevitably causes deterioration of the XPDC process. Moreover the conventionally used crystals, like diamond, can strongly absorb the photons and substantially suppress the XPDC efficiency[3].

In contrast, the XPDC using single electron, which is Kapitza-Dirac-like scattering, could have the advantage of being free of damage of the nonlinear crystal by the intense X-ray light and strong absorption of the photons inside of the crystal medium.

### S4.1 XPDC using nonlinear crystal

To obtain an order-of-magnitude estimation of the XPDC cross section by nonlinear crystalline medium, we use a semiclassical formalism to calculate nonlinear response functions for X-rays[5]. For electrons in the atom with density $\rho$ and velocity $\vec{v}$, we apply the equation of motion and continuity condition

$$\begin{aligned}\frac{\partial \vec{v}}{\partial t} + (\vec{v}\cdot\vec{\nabla})\vec{v} &= -\frac{e}{m}(\vec{E} + \frac{1}{c}\vec{v}\times\vec{B}) \\ \frac{\partial \rho}{\partial t} + \vec{\nabla}\cdot(\rho\vec{v}) &= 0.\end{aligned}\tag{S44}$$

Under perturbative expansion $\rho = \rho^{(0)} + \rho^{(1)} + \rho^{(2)} + \cdots$, $\vec{v} = \vec{v}^{(1)} + \vec{v}^{(2)} + \cdots$, and $\vec{J} = \vec{J}^{(1)} + \vec{J}^{(2)} + \cdots$, we obtain the second order nonlinear current for a general XPDC process $\omega_3 \to \omega_1 + \omega_2$

$$\begin{aligned}\vec{J}^{(2)}(\omega_3) &= \rho^{(0)}\vec{v}^{(2)} + \rho^{(1)}\vec{v}^{(1)} \\ &= \rho^{(0)}\frac{e^2}{m^2}\left[\frac{\vec{E}_1 \times (\vec{k}_2 \times \vec{E}_2)}{\omega_1\omega_2\omega_3} + i\frac{(\vec{E}_1 \cdot \vec{\nabla})\vec{E}_2}{\omega_1^2\omega_3}\right] \\ &\quad + \frac{ie^2}{m^2}\frac{(\vec{\nabla}\rho \cdot \vec{E}_2)\vec{E}_1}{\omega_1^2\omega_2} + \text{terms with interchanged index 1 and 2},\end{aligned} \quad (S45)$$

where the second term vanishes in the case $\vec{E}_1 \perp \vec{k}_2$, since $(\vec{E}_1 \cdot \vec{\nabla})\vec{E}_2 = (\vec{E}_1 \cdot \vec{k}_2)\vec{E}_2 = 0$.

We now consider the XPDC of a hard X-ray photon to a pair of X-ray and optical photons $\omega_{X3} \to \omega_{o1} + \omega_{X2}$. Physically this process is equivalent to the Thomson scattering of X-rays by an atom illuminated with an optical field, which induces Doppler-shifted sideband[6]. Of the several terms that occur in the second order current (Eq. S45) for the nonlinear response of an atom to applied electromagnetic fields of frequencies $\omega_{o1}$, $\omega_{X2}$, $\omega_{X3}$, only one is of importance to the present instance[6,7],

$$\vec{J}^{(2)}(\omega_{X3}) \simeq \frac{ie^2}{m^2}\frac{(\vec{\nabla}\rho \cdot \vec{E}_{X2})\vec{E}_{o1}}{\omega_{o1}^2\omega_{X2}}, \quad (S46)$$

where $\omega_{X3} = \omega_{o1} + \omega_{X2}$, and $\rho$ is the electron density of the crystal. Eq. S46 describes the Doppler-shifted reflection of $\vec{E}_2$ from electron that is driven by the optical field $\vec{E}_1$ and moving with velocity $\vec{v}_1$. For simplicity, we denote thereafter $\omega_{o1} = \omega_1$, $\omega_{X2} = \omega_2$, $\omega_{X3} = \omega_3$, the same notation applies for the momentum $k$ and electromagnetic fields $E$ and $B$. We expand the electron density in terms of reciprocal lattice vector $\vec{G}_m$ that $\rho(\vec{r}) = \sum_m \rho_m e^{i\vec{G}_m \cdot \vec{r}}$. The dominant nonlinear current is

$$\vec{J}^{(2)}(\omega_3) \simeq -\frac{e^2}{m^2}\sum_m \frac{(\rho_m \cdot \vec{e}_2)\vec{e}_1}{\omega_1^2\omega_2} e^{i[(\vec{k}_1 + \vec{k}_2 + \vec{G}_m) \cdot \vec{r} - (\omega_1 + \omega_2)t]}, \quad (S47)$$

where $\vec{e}_i$ is the polarization vector of the electric field $\vec{E}_i$. In the general case, the relation between the nonlinear response functions $R^{(n)}$ and the corresponding nonlinear susceptibilities $\chi^{(n)}$ is[5]

$$\chi^{(n)}(\omega;\omega_1,\omega_2,\cdots,\omega_n) = i^{1-n}\frac{c^n}{\omega_1\omega_2\omega_n}R^{(n)}(\omega;\omega_1,\omega_2,\cdots,\omega_n). \quad (S48)$$

For the quadratic process, the nonlinear current is

$$eJ_i^{(2)}(\omega_3) \propto R_{ijk}^{(2)}(\omega_3;\omega_1,\omega_2)A_j(\omega_1)A_k(\omega_2). \quad (S49)$$

Since $A(\omega) \sim \frac{cE}{\omega}$, by inspection the nonlinear response function is

$$R^{(2)} \propto \frac{e^3\rho}{m^2c^2\omega_1}, \quad (S50)$$

The nonlinear susceptibility is thus

$$\chi^2(\omega_3;\omega_1,\omega_2) \simeq \frac{e^3\rho}{m^2\omega_1^2\omega_2\omega_3} \stackrel{\text{a.u.}}{=} \frac{\alpha^2 r_e \rho}{k_1^2 k_2 k_3} \quad (S51)$$

in atomic units, where $\alpha$ is the fine structure constant and $r_e$ is the classical electron radius. The estimation of $\chi^{(2)}$ in Eq. S51 is consistent with the earlier work (see comment 28 in Ref. 8). Using Fermi's golden rule, we obtain the differential cross section for XPDC as[5]

$$\frac{d\sigma}{d\Omega}^{(2)} = \frac{\omega_3\omega_2^3\omega_1^3}{288\pi^3 c^7}\left|\chi^{(2)}\right|^2. \quad (S52)$$

Eqs. S51 and S53 give the scaling of cross section as $\propto 1/\omega_1$, which especially favors the low energy of the optical photon. For the instance given in the main text of XPDC $\omega_3(=3.1\,\text{keV}) \to \omega_2(=3096.9\,\text{eV}) + \omega_1(=3.1\,\text{eV})$ with diamond crystal, we obtain $\chi^{(2)} \sim 9.1 \times 10^{-12} \text{cm}^2 \text{StC}^{-1}$, where StC is the Gaussian unit of StatCoulomb. The XPDC cross section is

$$\frac{d\sigma}{d\Omega}^{(2)} \sim 1.9 \times 10^3 \, \text{fm}^2 = 19 \, \text{b}, \tag{S53}$$

which could be comparable to the cross section of Thomson scattering. Nevertheless, the low XPDC efficiency in actual experiments could be attributed to the radiation damage caused deterioration of phase matching condition and strong absorption of photons inside the crystal.

We also show here that the (quasi-)degenerate XPDC $\omega_X \to \omega_X/2 + \omega_X/2$ (Ref. 3) has cross section that is almost four orders of magnitude lower than the XPDC to X-ray and optical photons. In this case $\omega_3 = \omega$ and $\omega_1 = \omega_2 = \omega/2$ are both in the X-ray regime, the dominant nonlinear current is

$$\begin{aligned}\vec{J}^{(2)}(\omega_3) &\simeq \rho^{(0)}\frac{e^2}{m^2}\frac{\vec{E}_1 \times (\vec{k}_2 \times \vec{E}_2)}{\omega_1\omega_2\omega_3} \\ &\simeq \frac{e^2\rho^{(0)}\vec{k}_2 E_1 E_2}{m^2\omega_1\omega_2\omega_3}.\end{aligned} \tag{S54}$$

Similarly we obtain the nonlinear response function for (quasi-)degenerate XPDC

$$R^{(2)} \simeq \frac{\rho e^3}{2m^2c^3}, \tag{S55}$$

and the nonlinear susceptibility is in this case

$$\chi^{(2)} \simeq \frac{2e^3\rho}{m^2c\omega^3} \stackrel{\text{a.u.}}{=} \frac{\alpha^2 r_e \rho}{k^3}. \tag{S56}$$

Taken the experimental parameters in the earlier work[3], $\omega_3(=18\,\text{keV}) \to \omega_1(=9\,\text{keV}) + \omega_2(=9\,\text{keV})$ with diamond crystal, we obtain $\chi^{(2)} \sim 7.7 \times 10^{-20} \text{cm}^2 \text{StC}^{-1}$, and the XPDC cross section is

$$\frac{d\sigma}{d\Omega}^{(2)} \sim 4.7 \times 10^{-3} \, \text{b}, \tag{S57}$$

which is lower than the cross section in Eq. S53 by four orders of magnitude.

### S4.2 XPDC using single electron

As depicted in Fig. S3, the incident free electron plays the role of the nonlinear medium. A linearly polarized X-ray field ($\omega_{X3}$, $k_{X3}$) propagates from the right along the $x$ axis, while two seeding fields ($\omega_{o1}$, $k_{o1}$) and ($\omega_{X2}$, $k_{X2}$) propagate from the left along the $x$ axis, and $\omega_{o1} + \omega_{X2} = \omega_{X3}$, $k_{o1} + k_{X2} = k_{X3}$. The electric fields $E_{o1}$, $E_{X2}$, $E_{X3}$ are polarized along the $z$ axis. The figure eight motion of the electron in the fields of $E_{o1}$ and $E_{X2}$ created polarization along the $x$ axis $P_x^{(2)}(\omega_{X3}) = \chi_{xzz}^{(2)}(\omega_{X3}; \omega_{o1}, \omega_{X2})E_{o1,z}E_{X2,z}$. Meanwhile the X-ray field of $\omega_{X3}$ induces polarization $P_x^{(1)}(\omega_{X3}) = \chi_{xz}^{(1)}(\omega_{X3}; \omega_{X3})E_{X3}$. The two wave mixing of $P^{(2)}(\omega_{X3})$ and $P^{(2)}(\omega_{X3})$ induces the effective grating for diffractive scattering of the electron with Bragg angle $\sin\theta_B = \frac{k_{X3}}{p_i}$ (Fig. S3), where $p_i$ is the initial momentum of the electron[9]. From the perspective of nonlinear optics, this three color Kapitza-Dirac-like scattering is an XPDC process, in which the electron absorbs an X-ray photon of $\omega_{X3}$, converts it to a pair of photons with $\omega_{o1}$, $\omega_{X2}$ and changes the momentum by $2k_{X3}$ for momentum conservation. Since we use the fields of $\omega_{o1}$ and $\omega_{X2}$ as seeding fields, we could eventually identify the entangled photon pair from XPDC by coincidence with the deflected electron.

In the following we present the quantum-classical approach[10,11] to obtain the probability of the three color Kapitza-Dirac-like XPDC process $\omega_{X3} \to \omega_{o1} + \omega_{X2}$. For the scattering of the electron from the effective grating from the initial state $|\vec{p}_i\rangle$ to the final state $|\vec{p}_f\rangle = |\vec{p}_i + 2\vec{k}_{X3}\rangle$, we calculate the periodic stationary potential of the grating. As in Section S4.1, we denote thereafter $\omega_{o1} = \omega_1$, $\omega_{X2} = \omega_2$, $\omega_{X3} = \omega_3$, the same notation applies for the momentum $k$ and electromagnetic fields $E$ and $B$. For the mixed waves, the electric field $\vec{E} = (0,0,E(x,t))$ is

$$E(x,t) = E_1\cos(\omega_1 t - k_1 x) + E_2\cos(\omega_2 t - k_2 x) + E_3\cos(\omega_3 t + k_3 x). \tag{S58}$$

Faraday's law $\vec{\nabla}\vec{E} = -\frac{1}{c}\frac{\partial \vec{B}}{\partial t}$ gives the magnetic field $\vec{B} = (0, B(x,t), 0)$ with

$$B(x,t) = E_1\cos(\omega_1 - k_1 x) + E_2\cos(\omega_2 t - k_2 x) - E_3\cos(\omega_3 + k_3 x). \tag{S59}$$

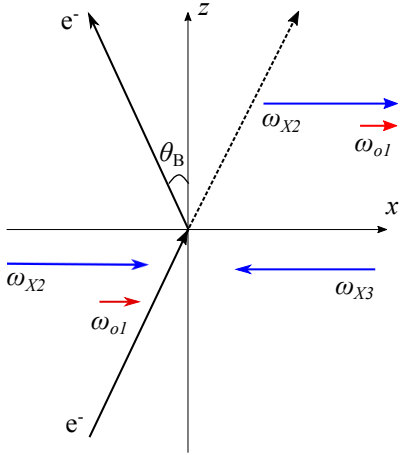

**Figure S3.** Sketch of the seeded Kapitza-Dirac-like XPDC process using single electron as the nonlinear medium. An X-ray photon of $\omega_{X3}$ is down converted to two photons of frequencies $\omega_{o1}$ and $\omega_{X2}$, as the electron is deflected with Bragg angle $\theta_B$ in coincidence. For the use of ghost diffraction, the down converted photon pair can then be spatially separated by a metal foil that reflects the optical photon. The collinear geometry of $k_{X3}$ and $(k_{o1}, k_{X2})$ can be loosened for more convenient experimental setup, provided the phase matching condition is satisfied.

The classical equations of motion of the electron in the electromagnetic field are

$$\ddot{x} = \frac{\dot{z}}{c} B(x,t) \tag{S60}$$

$$\ddot{z} = -E(x,t) - \frac{\dot{x}}{c} B(x,t). \tag{S61}$$

We separate the motion along $x$ and $z$ axis into fast and slow components, $x = x_f + x_s$, $z = z_f + z_s$, where the slow component corresponds to the electron motion across the laser focus, and the fast component corresponds to the oscillation of the electron in the light field. Applying Taylor expansion around $x_s$ and $z_s$, the equations of motion become

$$\ddot{x} \simeq \frac{\dot{z}}{c}\left[B(x_s,t) + x_f \frac{\partial B(x_s,t)}{\partial x_s}\right] \tag{S62}$$

$$\ddot{z} \simeq -E(x_s,t) - x_f \frac{\partial E(x_s,t)}{\partial x_s} - \frac{\dot{x}}{c}\left[B(x_s,t) + x_f \frac{\partial B(x_s,t)}{\partial x_s}\right]. \tag{S63}$$

Note that Eq. S61 is full derivative with respect to time, integrate it once we get

$$\dot{z} = v_z - 2\left[\frac{E_1}{\omega_1}\sin(\omega_1 t - k_1 x) - \frac{E_2}{\omega_2}\sin(\omega_2 t - k_2 x) - \frac{E_3}{\omega_3}\sin(\omega_3 t + k_3 x)\right]. \tag{S64}$$

Insert $\dot{z}$ of Eq. S64 into RHS of Eq. S62, and integrate twice over time to get the trajectory of $x(t)$. We inspect Eq. S62, keeping the right hand side (RHS) of Eq. S62 to the leading order of $\mathcal{O}(\dot{z}_s/c)$, the only possible terms in the bracket $[\cdots]$ that are independent of $t$ must rise from the product of second order terms in $x$ and the first order terms in $\frac{B(x_s,t)}{\partial x_s}$, because the $\omega_i t$ variables can then cancel with each other due to energy conservation condition $\omega_1 + \omega_2 = \omega_3$. We collect fast oscillating terms in $\ddot{x}$ of Eq. S60 that can lead to cancellation of the time variable, that

$$\begin{aligned}\ddot{x}_f = &-\frac{E_1 E_2}{\omega_1 c}\sin[(\omega_1+\omega_2)t - (k_1+k_2)x_s] - \frac{E_1 E_3}{\omega_1 c}\sin[(\omega_3-\omega_1)t + (k_3+k_1)x_s]\\ &-\frac{E_2 E_1}{\omega_2 c}\sin[(\omega_1+\omega_2)t - (k_1+k_2)x_s] - \frac{E_2 E_3}{\omega_2 c}\sin[(\omega_3-\omega_2)t + (k_3+k_2)x_s]\\ &-\frac{E_3 E_1}{\omega_3 c}\sin[(\omega_3-\omega_1)t + (k_3+k_1)x_s] - \frac{E_3 E_2}{\omega_3 c}\sin[(\omega_3-\omega_2)t + (k_3+k_2)x_s] + \mathcal{R},\end{aligned} \tag{S65}$$

where $\mathcal{R}$ stands for the residual terms that lead to non-stationary potentials. From Eq. S65 it is straightforward to

obtain

$$\begin{aligned}
x_f &= \frac{E_1 E_2}{\omega_1(\omega_1+\omega_2)^2 c}\sin[(\omega_1+\omega_2)t-(k_1+k_2)x_s] + \frac{E_1 E_3}{\omega_1(\omega_3-\omega_2)^2 c}\sin[(\omega_3-\omega_1)t+(k_3+k_1)x_s] \\
&+ \frac{E_2 E_1}{\omega_2(\omega_1+\omega_2)^2 c}\sin[(\omega_1+\omega_2)t-(k_1+k_2)x_s] + \frac{E_2 E_3}{\omega_2(\omega_3-\omega_2)^2 c}\sin[(\omega_3-\omega_2)t+(k_3+k_2)x_s] \\
&+ \frac{E_3 E_1}{\omega_3(\omega_3-\omega_1)^2 c}\sin[(\omega_3-\omega_1)t+(k_3+k_1)x_s] + \frac{E_3 E_2}{\omega_3(\omega_3-\omega_2)^2 c}\sin[(\omega_3-\omega_2)t+(k_3+k_2)x_s] \\
&+ \mathcal{R}.
\end{aligned} \tag{S66}$$

Insert Eq. S66 into the term $x_f \frac{\partial B(x_s,t)}{\partial x_s}$ of Eq. S62, and collect the stationary terms

$$\begin{aligned}
x_f \frac{\partial B(x_s,t)}{\partial x_s} &= x_f\left[\frac{\omega_1 E_1}{c}\sin(\omega_1 t - k_1 x_s) + \frac{\omega_2 E_2}{c}\sin(\omega_2 t - k_2 x_s) + \frac{\omega_3 E_3}{c}\sin(\omega_3 t + k_3 x_s)\right] \\
&= \frac{E_1 E_2 E_3}{c^2}\left(\frac{1}{\omega_1 \omega_3} + \frac{1}{\omega_1 \omega_2} + \frac{1}{\omega_2 \omega_3}\right)\cos[(k_1+k_2+k_3)x_s] + \mathcal{R}.
\end{aligned} \tag{S67}$$

Thus we have the equation of motion for the electron to travel across the light field,

$$\ddot{x}_s = \frac{z_s}{c^3} E_1 E_2 E_3 \left(\frac{1}{\omega_1 \omega_3} + \frac{1}{\omega_1 \omega_2} + \frac{1}{\omega_2 \omega_3}\right)\cos[(k_1+k_2+k_3)x_s]. \tag{S68}$$

Solving $\ddot{z}_s$ similarly and integrating once, we obtain

$$\dot{z}_s = v_z - \frac{E_1 E_2 E_3}{8\omega_3 c^2}\left(\frac{1}{\omega_1 \omega_3} + \frac{1}{\omega_1 \omega_2} + \frac{1}{\omega_2 \omega_3}\right)\sin[(k_1+k_2+k_3)x_s], \tag{S69}$$

where $v_z$ is the initial velocity of the electron along the $z$ axis. With Eq. S69 and S68 we obtain

$$\begin{aligned}
\ddot{x}_s &= \frac{v_z E_1 E_2 E_3}{c^3}\left(\frac{1}{\omega_1 \omega_3} + \frac{1}{\omega_1 \omega_2} + \frac{1}{\omega_2 \omega_3}\right)\cos[(k_1+k_2+k_3)x_s] \\
&- \frac{E_1^2 E_2^2 E_3^2}{16\omega_3 c^5}\left(\frac{1}{\omega_1 \omega_3} + \frac{1}{\omega_1 \omega_2} + \frac{1}{\omega_2 \omega_3}\right)^2 \sin[2(k_1+k_2+k_3)x_s].
\end{aligned} \tag{S70}$$

From the equation of motion for the scattering of the incident electron (Eq. S70), we can directly deduce the effective stationary periodic potential of the grating as

$$U_{\text{eff}} = \frac{v_z E_1 E_2 E_3}{2\omega_3 c^2}\left(\frac{1}{\omega_1 \omega_3} + \frac{1}{\omega_1 \omega_2} + \frac{1}{\omega_2 \omega_3}\right)\sin[(k_1+k_2+k_3)x_s], \tag{S71}$$

since the first term in Eq. S70 dominates. Using first order perturbation theory, the probability of elastic scattering of an incident free electron from $|\vec{p}_i\rangle$ to the final state $|\vec{p}_f\rangle = |\vec{p}_i + 2\vec{k}_3\rangle$ is

$$P(\omega_1,\omega_2,\omega_3) = \left|\frac{v_z E_1 E_2 E_3}{2c^2 \omega_1 \omega_2 \omega_3}\right|^2 \delta(E_{fi}), \tag{S72}$$

where $E_{fi} = p_f^2/2 - p_i^2/2$. The schematic setup in sketched in Fig. S3. To avoid the metal foil filter of being radiated by the intense X-ray field of $\omega_{X3}$, its collinear relation with the counterpropagating seeding fields with $\omega_{o1}$ and $\omega_{X2}$ can be loosened with respected to the phase matching condition.

## S5 Thomson and Rayleigh scattering in the optical and X-ray regime

In this section we elaborate the connection of Thomson and Rayleigh scattering in the X-ray and optical regime. We show that the frequency independent Thomson scattering stays intact for both regimes and the frequency dependent Rayleigh scattering corresponds to the dispersive corrections of form factor in the X-ray regime.

We apply the Hamiltonian for photon-electron interaction under velocity gauge [12]

$$\hat{H}_{I1} = -\frac{e}{m} \sum_{\vec{k},\vec{\varepsilon}} \sqrt{\frac{2\pi\hbar}{\omega V}} \int d^3r \hat{\Psi}^\dagger(\vec{r}) \left[ \vec{p} \cdot (\vec{\varepsilon}\hat{a}_{\vec{k}\vec{\varepsilon}} e^{i\vec{k}\cdot\vec{r}} + \vec{\varepsilon}^* \hat{a}^\dagger_{\vec{k}\vec{\varepsilon}} e^{-i\vec{k}\cdot\vec{r}}) \hat{\Psi}(\vec{r}) \right] \hat{\Psi}(\vec{r})$$

$$\hat{H}_{I2} = \frac{\hbar\pi e^2}{mV} \sum_{\vec{k},\vec{\varepsilon},\vec{k}',\vec{\varepsilon}'} \frac{1}{\sqrt{\omega\omega'}} \int d^3r \hat{\Psi}^\dagger(\vec{r}) (\vec{\varepsilon}^* \hat{a}^\dagger_{\vec{k}\vec{\varepsilon}} e^{-i\vec{k}\cdot\vec{r}} + \vec{\varepsilon}\hat{a}^\dagger_{\vec{k}\vec{\varepsilon}} e^{i\vec{k}\cdot\vec{r}})$$
$$\times (\vec{\varepsilon}'\hat{a}_{\vec{k}'\vec{\varepsilon}'} e^{i\vec{k}'\cdot\vec{r}} + \vec{\varepsilon}'^* \hat{a}^\dagger_{\vec{k}'\vec{\varepsilon}'} e^{-i\vec{k}'\cdot\vec{r}}) \hat{\Psi}(\vec{r}),  \quad (S73)$$

for the $\vec{p}\cdot\vec{A}$ and $A^2$ type interactions respectively. $\hat{\Psi}^{(\dagger)}$ and $\hat{a}^{(\dagger)}$ are the field operators for the electron and photon fields. $\vec{k}, \omega, \vec{\varepsilon}$ are the momentum, frequency and polarization vector of the photon. $V$ is the quantization volume.

The $A^2$ type interaction from $\hat{H}_{I2}$ describes the Thomson scattering [12]. Assume the initial and final states are

$$|I\rangle = |\Psi_a\rangle |N\rangle$$
$$|F\rangle = \frac{1}{\sqrt{N}} |\Psi_a\rangle \hat{a}^\dagger_{\vec{k}_f \vec{\varepsilon}_f} |N-1\rangle, \quad (S74)$$

where $|N\rangle$ denotes the Fock state of the photon field and $|\Psi_a\rangle$ is the electronic eigenstate of the molecule. Note in the summation $\sum_{\vec{k},\vec{\varepsilon},\vec{k}',\vec{\varepsilon}'} (\vec{\varepsilon}^* \hat{a}^\dagger_{\vec{k}\vec{\varepsilon}} e^{-i\vec{k}\cdot\vec{r}} + \vec{\varepsilon}\hat{a}^\dagger_{\vec{k}\vec{\varepsilon}} e^{i\vec{k}\cdot\vec{r}})(\vec{\varepsilon}'\hat{a}_{\vec{k}'\vec{\varepsilon}'} e^{i\vec{k}'\cdot\vec{r}} + \vec{\varepsilon}'^* \hat{a}^\dagger_{\vec{k}'\vec{\varepsilon}'} e^{-i\vec{k}'\cdot\vec{r}})$ for $\vec{k} = \vec{k}_i, \vec{k}_f, \vec{k}' = \vec{k}'_i, \vec{k}'_f$, the terms of nonzero contribution to the transition is $[\hat{a}_{\vec{k}_i\vec{\varepsilon}_i} \hat{a}^\dagger_{\vec{k}_f\vec{\varepsilon}_f} + \hat{a}^\dagger_{\vec{k}_f\vec{\varepsilon}_f} \hat{a}_{\vec{k}_i\vec{\varepsilon}_i}] e^{i(\vec{k}_i - \vec{k}_f)\cdot\vec{r}}$. The transition matrix element can be calculated as

$$M_{I2} = \frac{\pi\hbar e^2}{mV} \frac{(\vec{\varepsilon}_i \cdot \vec{\varepsilon}_f^*)}{\sqrt{\omega_i \omega_f}} \langle N-1| \hat{a}_{\vec{k}_f\vec{\varepsilon}_f} [2\hat{a}^\dagger_{\vec{k}_f\vec{\varepsilon}_f} \hat{a}_{\vec{k}_i\vec{\varepsilon}_i} + \delta_{\vec{k}_f \vec{k}_i} \delta_{\vec{\varepsilon}_f \vec{\varepsilon}_i}] |N\rangle \int d^3r \langle \Psi_a| \hat{\Psi}^\dagger(\vec{r}) e^{i(\vec{k}_i - \vec{k}_f)\cdot\vec{r}} \hat{\Psi}|\Psi_a\rangle$$
$$= \frac{2\pi\hbar e^2}{mV} \frac{\vec{\varepsilon}_i \cdot \vec{\varepsilon}_f^*}{\sqrt{\omega_i \omega_f}} f^{(0)}(\vec{Q}), \quad (S75)$$

where the form factor is

$$f^{(0)}(\vec{Q}) = \int d^3r \langle \Psi_a| \hat{\Psi}^\dagger(\vec{r}) \hat{\Psi}(\vec{r}) |\Psi_a\rangle e^{i\vec{Q}\cdot\vec{r}}$$
$$= \int d^3r \rho(\vec{r}) e^{i\vec{Q}\cdot\vec{r}} \quad (S76)$$

for $\vec{Q} = \vec{k}_i - \vec{k}_f$. Fermi's golden rule gives the elastic Thomson cross section for $\omega = \omega_i = \omega_f$

$$Jd\sigma = \frac{2\pi}{\hbar} \sum_{\vec{k}_f} |M_{I2}|^2 \delta(\hbar(\omega_f - \omega_i))$$
$$= \frac{2\pi}{\hbar} \frac{V}{(2\pi)^3} d\Omega \int dk_f k_f^2 |M_{I2}|^2 \frac{1}{\hbar} \delta(\omega_f - \omega_i)$$
$$= \frac{2\pi}{\hbar} \frac{V}{(2\pi)^3} \frac{1}{c^3} \frac{1}{\hbar} d\Omega \int d\omega_f \omega_f^2 |M_{I2}|^2 \delta(\omega_f - \omega_i)$$
$$= \frac{2\pi}{\hbar} \frac{V}{(2\pi)^3} \frac{1}{c^3} \frac{\omega^2}{\hbar} d\Omega |M_{I2}|^2$$
$$\Rightarrow \frac{d\sigma}{d\Omega} = \frac{V^2 \omega^2}{4\pi^2 c^4 \hbar^2} |M_{I2}|^2$$
$$= \frac{e^4}{m^2 c^4} |f^{(0)}(\vec{Q})|^2 \sum_{\vec{\varepsilon}_f} |\vec{\varepsilon}_i \cdot \vec{\varepsilon}_f^*|^2$$
$$= r_e^2 |f^{(0)}(\vec{Q})|^2 \frac{1}{2}(1 + \cos^2\theta), \quad (S77)$$

where $J = c/V$ is the normalized photon flux per unit area and unit time. Thus the Thomson scattering remains the same for optical and X-ray photons because it is independent of the photon frequency.

Next we show the equivalence of Rayleigh scattering in optical regime to the dispersive correction of form factor $f'(\omega)+if''(\omega)$ in X-ray regime, such that the complete form factor is $f(\vec{Q},\omega) = f^{(0)}(\vec{Q}) + f'(\omega) + if''(\omega)$. In the X-ray regime, the dispersive effects are used as a solution to the phase problem in crystallography[13,14].

The Rayleigh scattering is described by the second order interaction of $\vec{p} \cdot \vec{A}$ type. The corresponding transition matrix element is

$$\begin{aligned}
M_{I1} &= \frac{2\pi\hbar e^2}{mV} \frac{1}{\sqrt{\omega_i \omega_f}} \\
&\quad \times \sum_r \Bigg\{ \langle \Psi_a | \int d^3 r \hat{\Psi}(\vec{r}) e^{-i\vec{k}_f \cdot \vec{r}} \vec{p} \cdot \vec{\varepsilon}_f^* \hat{\Psi}(\vec{r}) | \Psi_r \rangle \frac{1}{E_a - E_r + \hbar\omega_i + i\varepsilon} \langle \Psi_r | \int d^3 r \hat{\Psi}(\vec{r}) e^{i\vec{k}_i \cdot \vec{r}} \vec{p} \cdot \vec{\varepsilon}_i \hat{\Psi}(\vec{r}) | \Psi_a \rangle \\
&\quad + \langle \Psi_a | \int d^3 r \hat{\Psi}(\vec{r}) e^{i\vec{k}_i \cdot \vec{r}} \vec{p} \cdot \vec{\varepsilon}_i \hat{\Psi}(\vec{r}) | \Psi_r \rangle \frac{1}{E_a - E_r - \hbar\omega_i - i\varepsilon} \langle \Psi_r | \int d^3 r \hat{\Psi}(\vec{r}) e^{i\vec{k}_i \cdot \vec{r}} \vec{p} \cdot \vec{\varepsilon}_i \hat{\Psi}(\vec{r}) | \Psi_a \rangle \Bigg\} \\
&= \frac{2\pi\hbar e^2}{m\omega V} \left[ f'(\omega) + if''(\omega) \right] ,
\end{aligned} \tag{S78}$$

where $|\Psi_r\rangle$ are the virtual states for the second order transition. The dispersive corrections follow the standard definition in X-ray physics by decomposing the terms in the bracket of Eq. S78 using the identity $\frac{1}{x+i\varepsilon} = \text{Pr}\frac{1}{x} - i\pi\delta(x)$ [12,14].

It can be easily shown that the matrix element of $\vec{p} \cdot \vec{A}$ interaction (Eq. S78) relates to the classical Rayleigh scattering cross section in the optical regime, which has the $(\omega/\omega_0)^4$ dependence on the photon frequency. For this purpose, we reduce the model to two-level system such that the classical *natural frequency* $\omega_0$ can be defined through $\hbar\omega_0 = E_r - E_a$, and assume dipole approximation $e^{i\vec{k}\cdot\vec{r}} \simeq 1$. Since $\omega \ll omega_0$, we make expansion $\frac{1}{\omega_0+\omega} + \frac{1}{\omega_0-\omega} = \frac{1}{\omega_0} + \frac{\omega^2}{\omega_0^3} + \cdots$, and insert the expansion into Eq. S78, the matrix element reduces to

$$\begin{aligned}
M_{I1} &= -\frac{2\pi e^2}{mV} \frac{1}{\omega} \Bigg\{ \langle \Psi_a | \int d^3 r \hat{\Psi}(\vec{r}) \vec{p} \cdot \vec{\varepsilon}_f^* \hat{\Psi}(\vec{r}) | \Psi_r \rangle \frac{1}{\omega_0} \langle \Psi_r | \int d^3 r \hat{\Psi}(\vec{r}) \vec{p} \cdot \vec{\varepsilon}_i \hat{\Psi}(\vec{r}) | \Psi_a \rangle \\
&\quad + \langle \Psi_a | \int d^3 r \hat{\Psi}(\vec{r}) \vec{p} \cdot \vec{\varepsilon}_i \hat{\Psi}(\vec{r}) | \Psi_r \rangle \frac{1}{\omega_0} \langle \Psi_r | \int d^3 r \hat{\Psi}(\vec{r}) \vec{p} \cdot \vec{\varepsilon}_i \hat{\Psi}(\vec{r}) | \Psi_a \rangle \Bigg\} \\
&\quad - \frac{2\pi e^2}{mV} \frac{\omega}{\omega_0^2} \Bigg\{ \langle \Psi_a | \int d^3 r \hat{\Psi}(\vec{r}) \vec{p} \cdot \vec{\varepsilon}_f^* \hat{\Psi}(\vec{r}) | \Psi_r \rangle \frac{1}{\omega_0} \langle \Psi_r | \int d^3 r \hat{\Psi}(\vec{r}) \vec{p} \cdot \vec{\varepsilon}_i \hat{\Psi}(\vec{r}) | \Psi_a \rangle \\
&\quad + \langle \Psi_a | \int d^3 r \hat{\Psi}(\vec{r}) \vec{p} \cdot \vec{\varepsilon}_i \hat{\Psi}(\vec{r}) | \Psi_r \rangle \frac{1}{\omega_0} \langle \Psi_r | \int d^3 r \hat{\Psi}(\vec{r}) \vec{p} \cdot \vec{\varepsilon}_i \hat{\Psi}(\vec{r}) | \Psi_a \rangle \Bigg\} + \cdots ,
\end{aligned} \tag{S79}$$

Define the polarizibility tensor[15]

$$\ddot{\alpha} \stackrel{\text{def}}{=} \frac{2}{\hbar} \sum_r \frac{\langle \Psi_a | \vec{p} | \Psi_r \rangle \langle \Psi_r | \vec{p} | \Psi_a \rangle}{\omega_a - \omega_r} , \tag{S80}$$

we can write Eq. S79 in a simple form

$$M_{I1} \simeq \frac{2\pi\hbar e^2}{mV\omega} \vec{\varepsilon}_i \ddot{\alpha} \vec{\varepsilon}_f^* + \frac{2\pi\hbar e^2 \omega}{mV\omega_0^2} \vec{\varepsilon}_i \ddot{\alpha} \vec{\varepsilon}_f^* + \cdots . \tag{S81}$$

We take constant polarization approximation that $\ddot{\alpha} = \alpha_0 \ddot{\mathbb{I}}$, and Eq. S81 gives

$$M_{I1} \simeq \frac{2\pi\hbar e^2}{mV\omega} (\vec{\varepsilon}_i \cdot \vec{\varepsilon}_f^*) + \frac{2\pi\hbar e^2 \omega}{mV\omega_0^2} (\vec{\varepsilon}_i \cdot \vec{\varepsilon}_f^*) + \cdots . \tag{S82}$$

Without considering interference between the two terms in Eq. S82, it gives the Rayleigh scattering cross section of second order $\vec{p} \cdot \vec{A}$ interaction in the optical regime, and recall its form in the X-ray regime in Eq. S78,

$$\begin{aligned}
\frac{d\sigma}{d\Omega} &= \frac{V^2 \omega^2}{4\pi^2 c^4 \hbar^2} |M_{I1}|^2 \\
&= r_e^2 \left[ \alpha_0^2 \left( 1 + \frac{\omega_4}{\omega_0^4} \right) \right] \frac{1}{2}(1+\cos^2\theta) \\
&= r_e^2 |f'(\omega) + if''(\omega)|^2 .
\end{aligned} \tag{S83}$$

From Eq. S83 we could conceive the usage of Rayleigh scattering in the entangled optical X-ray ghost diffraction, as it is applied to the phase problem of X-ray crystallography.

## References

1. Strekalov, A., Sergienko, A. V., Klyshko, D. N. & Shih, Y. H. Observation of two-photon "ghost" interference and diffraction. *Phys. Rev. Lett.* **74**, 3600 (1995).

2. Rubin, M. H. & Shih, Y. H. Resolution of ghost imaging for nondegenerate spontaneous parameteric down-conversion. *Phys. Rev. A* **78**, 033836 (2008).

3. Shwartz, S. *et al.* X-ray parametric down-conversion in the langevin regime. *Phys. Rev. Lett.* **109**, 013602 (2012).

4. Rubin, M. H. Transverse correlation in optical spontaneous parametric down-conversion. *Phys. Rev. A* **54**, 5349 (1996).

5. Zambianchi, P. Nonlinear response functions for X-Ray laser pulses. In *Nonlinear Optics, Quantum Optics, and Ultrafast Phenomena with X-Rays*, 287–302 (Springer Verlag, 2003).

6. Freund, I. & Levine, B. F. Optically modulated X-Ray diffraction. *Phys. Rev. Lett.* **25**, 1241–1245 (1970).

7. Freund, I. Nonlinear X-ray diffraction. Determination of valence electron charge distributions. *Chem. Phys. Lett.* **12**, 583 (1972).

8. Barbiellini, B., Joly, Y. & Tamasaku, K. Explaining the X-ray nonlinear susceptibility of diamond and silicon near absorption edges. *Phys. Rev. B* **92**, 155119 (2015).

9. Boyd, R. *Nonlinear Optics* (Academic Press, 1992).

10. Smirnova, O., Freimund, D. L., Batelaan, H. & Ivanov, M. Kapitza-Dirac diffraction without standing waves: Diffraction without a grating? *Phys. Rev. Lett.* **92**, 223601 (2004).

11. Fedorov, M. V. *Electrons in a Strong Laser Field* (Nauka, Moscow, 1991).

12. Santra, R. Concepts in X-ray physics. *J. Phys. B: Atomic, Molecular and Optical Physics* **42**, 023001 (2008).

13. Son, S.-K., Chapman, H. N. & Santra, R. Multiwavelength anomalous diffraction at high X-ray intensity. *Phys. Rev. Lett.* **107**, 218102 (2011).

14. Als-Nielsen, J. & McMorrow, D. *Elements of modern X-ray physics* (John Wiley & Sons, 2011).

15. Long, D. A. Quantum mechanical theory of Rayleigh and Raman scattering. In *The Raman Effect: A Unified Treatment of the Theory of Raman Scattering by Molecules*, 49–84 (John Wiley & Sons, 2002).